\documentclass{article}
\usepackage{epsfig}
\usepackage{subfigure}

\begin{document}

\title{\bf First Physical Results from SND Detector at VEPP-2M }
\author{
M.N.Achasov,
M.G.Bek,
K.I.Beloborodov,
A.V.Berdyugin, \\
A.V.Bozhenok,
A.D.Bukin,
D.A.Bukin,
S.V.Burdin, \\
V.V.Danilov,
T.V.Dimova,
S.I.Dolinsky,
V.P.Druzhinin, \\
M.S.Dubrovin,
I.A.Gaponenko,
V.B.Golubev,
V.N.Ivanchenko, \\
P.I.Ivanov,
I.A.Koop,
A.A.Korol,
M.S.Korostelev, \\
S.V.Koshuba,
A.P.Lysenko,
A.A.Mamutkin,
I.N.Nesterenko, \\
E.V.Pakhtusova,
E.A.Perevedentsev,
A.A.Polunin,
E.G.Pozdeev, \\
V.I.Ptitsyn,
E.E.Pyata,
A.A.Salnikov,
A.V.Savchkov, \\
S.I.Serednyakov,
V.V.Shary,
Yu.M.Shatunov,
V.A.Sidorov, \\
Z.K.Silagadze,
A.N.Skrinsky,
Yu.V.Usov,
A.A.Valishev, \\
A.V.Varganov,
A.V.Vasiljev,
Yu.S.Velikzhanin.\\ \\
\it The Budker Institute of Nuclear Physics,
630090, Novosibirsk, Russia
}

\date{September 1997}

\maketitle

\begin{abstract}
        The paper describes experiments with the SND detector at
VEPP-2M collider, carried out during the period from October 1995
until June 1997. The total integrated luminosity of $6.4~pb^{-1}$
was collected in the energy range $2E=0.4 \div 1.4~GeV$ (MHAD97
experiment), corresponding
to $4\cdot 10^5$ $\mu^+ \mu^-$-pairs produced. Preliminary results
of the 1996 $\phi$-meson experiment (FI96) are presented. The total number
of $\phi$-mesons produced is $4\cdot 10^6$. New data on rare decays
of $\phi$ and $\eta (550)$ mesons, in particular \\
\hspace*{10mm} $B(\phi \to \eta \gamma ) = (1.30\pm 0.06\pm 0.07 )\ \%$, \\
\hspace*{10mm} $B(\phi\to\pi^o \pi^o \gamma )\ = (1.1 \pm 0.2)\cdot 10^{-4}$,
$(M_{\pi^o\pi^o} > 800 MeV)$,\\
\hspace*{10mm} $B(\phi\to f^o \gamma)\ = (4.7\pm 1.0) \cdot 10^{-4}$,\\
\hspace*{10mm} $B(\phi\to\eta \pi^o \gamma )\ = (1.3 \pm 0.5) \cdot 10^{-4}$,\\
\hspace*{10mm} $B(\phi\to\eta '\gamma )\ < 1.7 \cdot 10^{-4}$,\\
\hspace*{10mm} $B(\phi\to 2\pi^o )\ < 6 \cdot 10^{-4}$\\
were obtained.

\end{abstract}

\newpage

\section{ The MHAD97 experiment }

\begin{figure}[htb]

\vspace{3cm}

  \begin{center}
    \mbox{\epsfig{figure=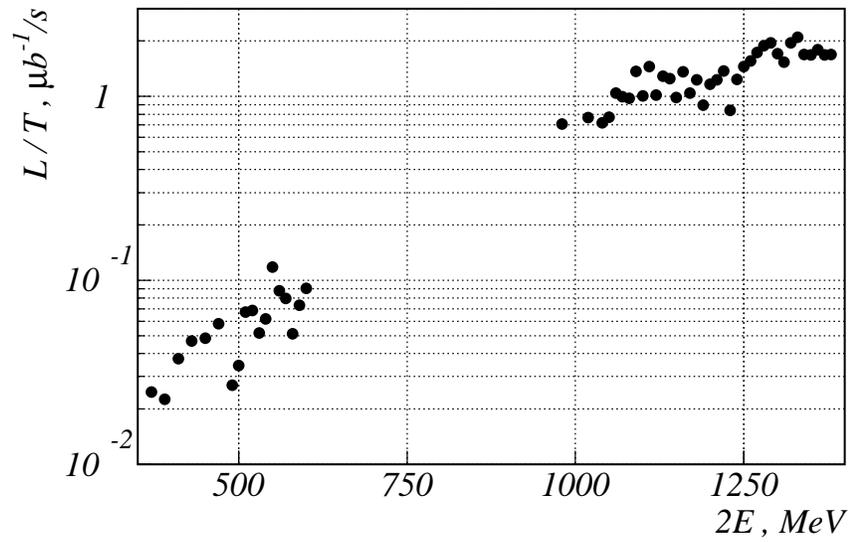,%
                        width=\textwidth}}
  \end{center}
\caption{Average luminosity (integrated luminosity at each
energy point divided by corresponding live time of data taking)
in the experiments with SND detector as a function of the beam
energy.}
\label{LUM}
\end{figure}

In the end of 1996 the experiments, started in 1995 with SND detector
\cite{SND1,SND2}, were continued. In the period from October until November
the energy range $2E$ from 600 down to 370~$MeV$ was scanned. The total
integrated luminosity of $80~nb^{-1}$ was collected, which corresponds
to approximately $3\cdot 10^4 \mu^+ \mu^-$ and $10^4 \pi^+ \pi^-$ events.
The goal of the experiment was to collect data needed to improve
accuracy of  calculations of the contribution of
this energy region into anomalous magnetic moment of muon. The data were
taken in parallel with the CMD--2 \cite{CMD2} detector. The energy dependence of 
VEPP-2M average luminosity during this experiment is shown in Fig.~\ref{LUM}.

In the beginning of January 1997 the VEPP-2M energy was set to
approximately $2E=M_{\phi}$ and data were collected corresponding to
integrated luminosity of about $200~nb^{-1}$ or $0.7\cdot 10^6$
$\phi$-mesons produced.

\begin{figure}[htb]
  \begin{center}
    \mbox{\epsfig{figure=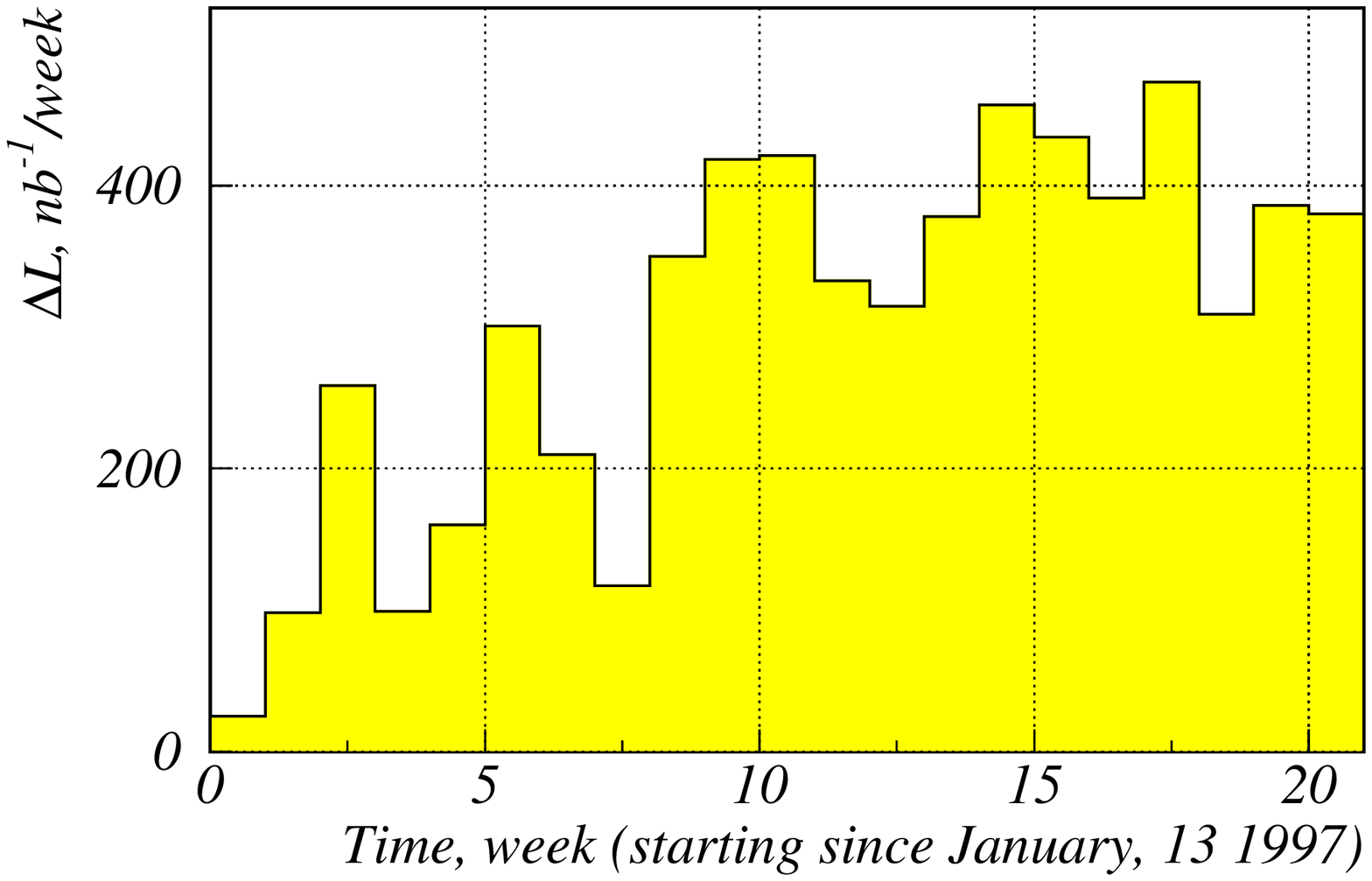,%
                    width=\textwidth}}
  \end{center} 
\caption{ SND total integrated luminosity
accumulation schedule in the period from 01/13/1997
to 06/13/1997.}
\label{LUM_WEEK}
	
  \begin{center}
    \mbox{\epsfig{figure=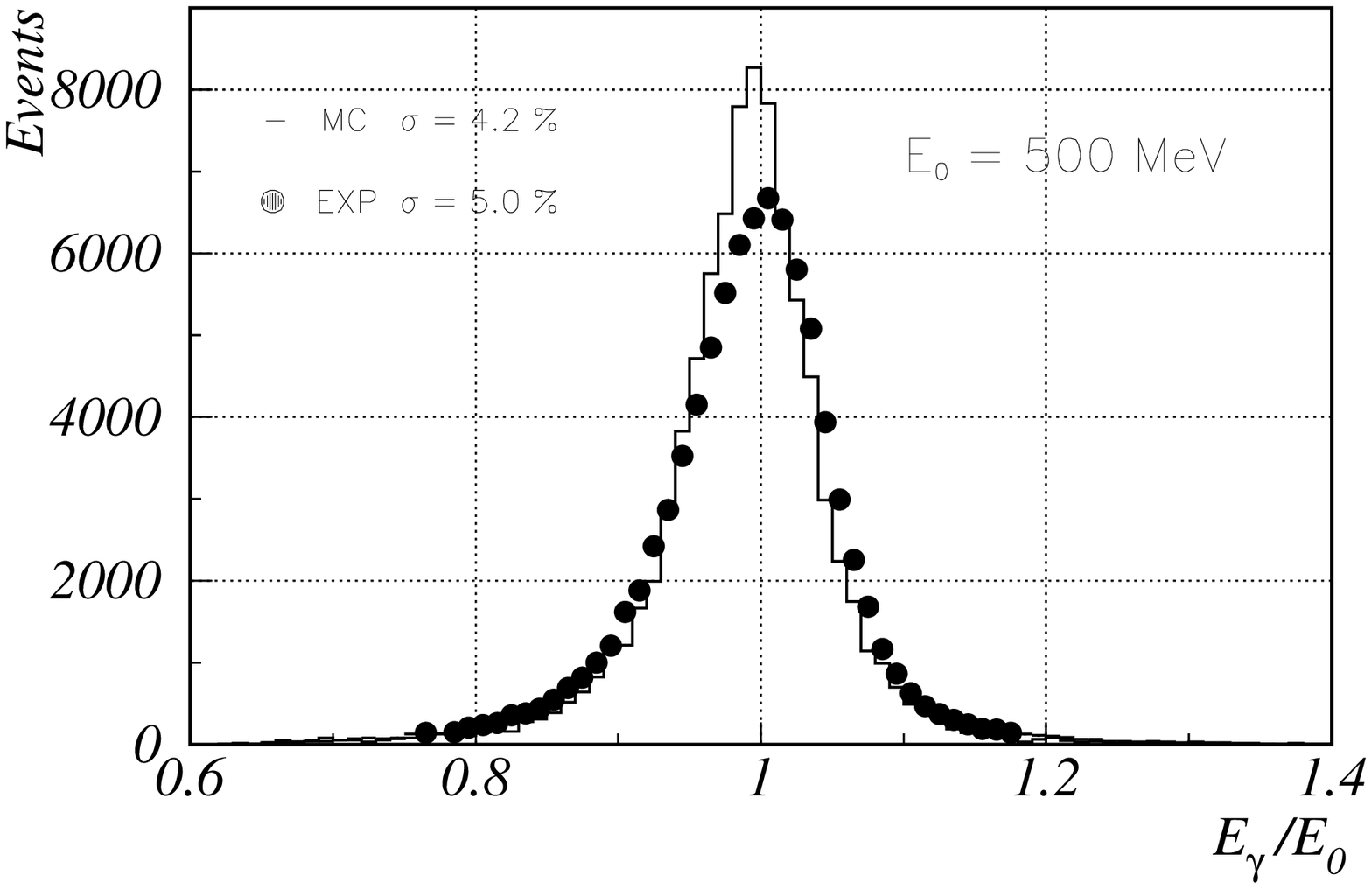,%
                         width=\textwidth}}
 \end{center} 
\caption{Distribution of $e^+e^- \to \gamma \gamma $ events over
normalized energy deposition in the calorimeter. Points ---
experimental data, $\sigma _{exp.}=5.0\%$; histogram ---
simulation, $\sigma _{M.C.}=4.2\%$.}
\label{ETONEE}
\end{figure}

From the end of January until June 1997, in parallel with the CMD--~2
detector, the MHAD97 experiment in the energy range $2E$ from
980 to 1380~$MeV$ was carried out. Two scans of this energy region
were performed: one -- upwards,
and another -- downwards, with a step of $\Delta (2E) = 10~MeV$.
The total integrated luminosity collected was $6.3~pb^{-1}$ at an average
luminosity of $1.3\cdot 10^{30}~cm^{-2} s^{-1}$ (Fig.~\ref{LUM}),
and the total number of recorded events of $ 10^8$, from which about
$3.6\cdot 10^5$ were $\mu^+ \mu^-$- pairs, and $10^5$  $\pi^+ \pi^-$- pairs.
In Fig. ~\ref{LUM_WEEK} the weekly schedule of data taking is shown.
The mean event recording rate was about $\sim 23~Hz$ and average
live time $\sim 72~\%$. The raw events were stored on thirty 8~mm 4~$GB$ tapes.

The main goal of this experiment was a thorough measurement of different
hadronic production cross sections, including
$e^+e^- \to 2\pi , 3\pi ,4\pi , \omega \pi ,$ $ K \bar{K} , K \bar{K} \pi ,
\dots $. New measurements can facilitate more precise tests of different theoretical
models (VDM, CVC, ...), check if there are sizable contributions from radial
excitations $\rho ', \omega ', \phi '$, and determine the contribution
of this energy region into muon anomalous magnetic moment.


\section{ Detector performance }
The SND detector was described in detail elsewhere in \cite{SND,SND1,SND2}.
Simultaneously with data taking the work on improvement of detector
parameters was performed. One of the most important parameters is the energy
resolution of the calorimeter for electrons and photons. Primary calibration
of the calorimeter was done using cosmic muons, and final calibration --
using Bhabha events. Recently, progress in understanding of
factors limiting the detector resolution (Fig.\ref{ETONEE})   was
achieved. It turned out, that nonuniformity of the light collection
efficiency in the NaI(Tl) crystals of the inner calorimeter layer
contributed significantly into  calorimeter response.
When this effect was included into simulation program, the disagreement
between measured ($\sigma _E /E_\gamma = 5.5\%$ at $E_\gamma =500~MeV$)
and simulated resolution ($\sigma _E /E_\gamma = 4.2\%$)
was significantly reduced. The next step in improvement of calorimeter
response was implementation of an absolute calibration procedure based
on Bhabha scattering events. With such a calibration the energy resolution
for electrons and photons was improved by approximately 10\%. Such a modest
improvement with respect to "cosmic" calibration \cite{CALIB_CALORIM}
shows relatively high precision of the cosmic calibration procedure.

Another important parameter of the calorimeter is an invariant mass
resolution $\sigma _m/m$ for particles decaying into photons, which
depends on a particle energy and mass. Table~\ref{DMTOM} shows
values of  $\sigma _m/m$ for $\pi^o$, $\eta$, $K_S$
ט $\omega$- mesons, extracted from FI96 experimental data  \cite{SND2}.

\begin{table}[!htb]
\caption{ Invariant mass resolution ($\sigma _m/m$)
for particles decaying into photons. }
\label{DMTOM}
\footnotesize
\begin{center}

\begin{tabular}{||c|c|c|c|c|c||}

\hline
Particle   & $\pi^o$   & $\eta$  & $\eta$  & $K_S$  & $\omega$  \\
\hline
Energy, $MeV$ &  519 & 657 &  657  & 510 & 801 \\
Process & $\phi \to \pi^o \gamma $
        & $\phi \to \eta  \gamma $
        & $\phi \to \eta  \gamma $
        & $\phi \to K_S K_L $
        & $\phi \to \omega \pi^o $
        \\
Decay mode   & $\pi^o \to \gamma \gamma$
        & $\eta  \to \gamma \gamma$
        & $\eta \to 3\pi^o$
        & $K_S \to 2\pi^o$
        & $\omega \to \pi^o \gamma $ \\
$\sigma _m/m$, MC.,    \% & 8.1 & 2.7 & 2.9 & 4.4 & 1.8 \\
$\sigma _m/m$, exp.,   \% & 10 & 2.9 & 3.5 & 5.2 & 2.4 \\
\hline
\end{tabular}
\end{center}
\end{table}

Systematic
discrepancy between measured and simulated resolutions, 
seen in the Table~\ref{DMTOM}, is still
under investigation. For example, in Fig.~\ref{MPI0G} the $\omega$- meson
invariant mass spectrum
in the decay channel $ \omega \to \pi^o \gamma $ is shown. It was obtained
during the study of the process
$e^+e^- \to \omega \pi^o \to \pi^o \pi^o \gamma $.

\begin{figure}[htb]
  \begin{center}
    \mbox{\epsfig{figure=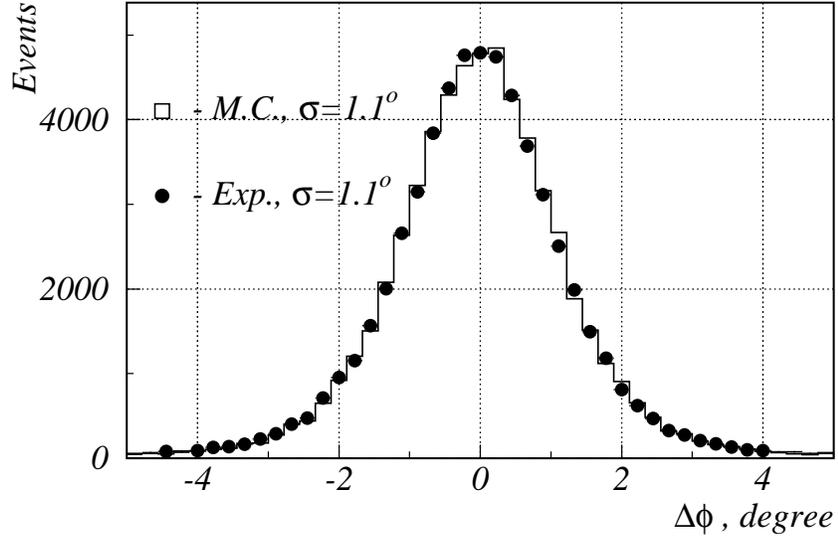,%
                   width=\textwidth}}
  \end{center}
\caption{ Acollinearity angle distribution
in azimuth plane in the  $e^+e^- \to e^+e^-$ events.}
\label{DPHI}
\end{figure}

\begin{figure}[htb]
  \begin{center}
    \mbox{\epsfig{figure=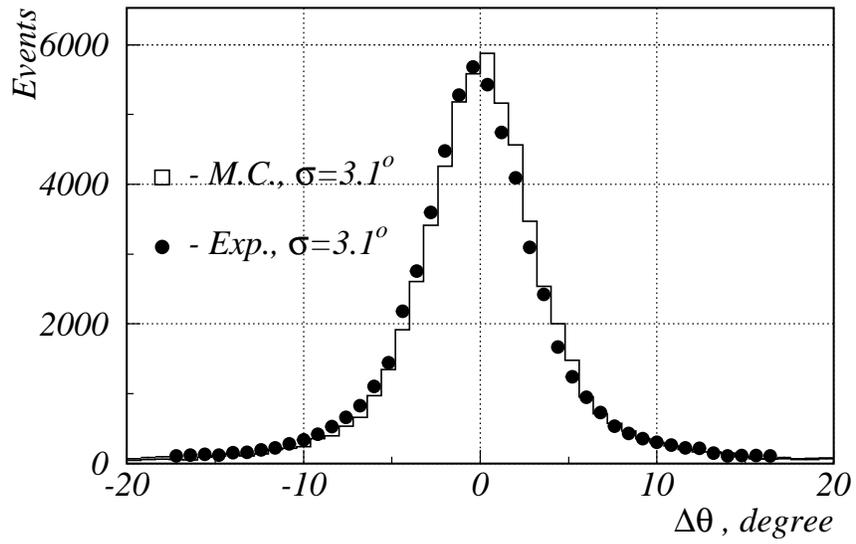,%
                   width=\textwidth}}
  \end{center} 
\caption{ Polar acollinearity angle distribution
in the $e^+e^- \to e^+e^-$ events.}
\label{DTHETA}
\end{figure}

\begin{figure}[htb]
  \begin{center}
    \mbox{\epsfig{figure=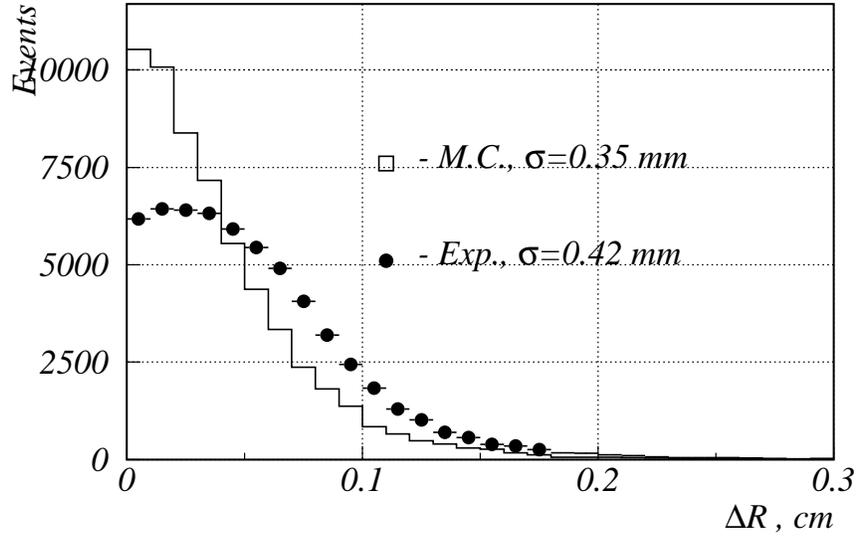,%
                   width=\textwidth}}
  \end{center} 
\caption{ Distribution over the distance from a track to the beam
axis $\Delta R$ in the $e^+e^- \to e^+e^-$ events.}
\label{DR}
\end{figure}

Before the beginning of MHAD97 experiment the performance of
the drift chamber system was improved, using for calibration copiously
produced collinear $e^+e^-$- pairs from Bhabha scattering process.
The spatial resolution in transverse direction,
determined by the accuracy of drift time measurements was equal to
180 microns ($\sigma$) on average;
the resolution in longitudinal direction was 3.2~mm ($\sigma$) for
charge division, and 1.5~mm for cathode strip readout.
In Fig.~\ref{DPHI}, \ref{DTHETA} the experimental and simulated
distributions over
$\Delta \varphi$, $\Delta \vartheta$ angles in the Bhabha
scattering events are shown. The widths of the distributions are
determined not only by spatial resolution of the drift chambers,
but also by higher order QED corrections to Bhabha scattering.
The intrinsic angular
resolution of the drift chambers is $\sigma _\phi = 0.7^o$ in
azimuth and $\sigma _\vartheta = 2.2^o$ in polar plane. The performance of
drift chambers can also be described in terms of impact parameter
$\Delta R$ -- measured distance between the track of a charged particle
and beam axis, projected onto a plane, perpendicular to the beam
(Fig.~\ref{DR}). In the infinite spatial resolution limit the average
$\Delta R$ value must be equal to the transverse size of the beam
$\sigma _x =10$ microns, $\sigma _r =200$ microns. For the Bhabha
scattering events $\sigma _{\Delta R} \simeq 0.4~mm$ was observed.

The luminosity was monitored during the experiment using
the processes of double brems\-strahlung (by an external small-angle
monitor), Bhabha scattering, and 2$\gamma$\-annihilation. Listed below are the
selection criteria for events of the two latter processes, used for
luminosity monitoring.

{\bf For the $e^+e^- \to e^+e^-$} process:
number of particles $\leq 4$;
number of charged particles $\geq 2$;
distance between tracks and beam collision point
$\Delta R \leq 1~cm$,
$\Delta Z \leq 10~cm$;
acollinearity angle
$\Delta \varphi\leq 10^o$, $\Delta \vartheta\leq 25^o$;
polar angles of both tracks are within
$27^מ \leq \vartheta \leq 153^o$ interval;
energy deposition relative to the beam energy
$0.6 \leq \Delta E/E \leq 1.5$ for both particles;
two-charged-particles trigger must be produced in the event;

{\bf For the $e^+e^- \to \gamma \gamma$ process}:
number of particles $\leq 4$;
number of neutral particles $\geq 2$;
acollinearity angle between $\gamma$- quanta
$\Delta \varphi\leq 10^o$, $\Delta \vartheta\leq 25^o$;
polar angles of both photons are within
$27^מ \leq \vartheta \leq 153^o$ interval;
energy deposition relative to the beam energy
$0.6 \leq \Delta E/E \leq 1.5$ for both photons;
Two-neutral-particles trigger must be produced in the event.

\begin{figure}[htb]
  \begin{center}
    \mbox{\epsfig{figure=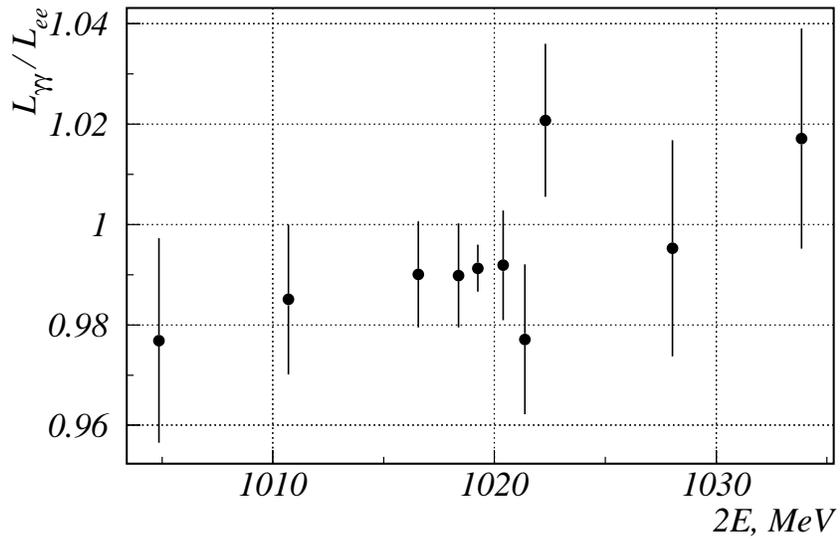,%
                         width=\textwidth}}
  \end{center} 
\caption{ Ratio of the luminosities $L_{\gamma \gamma} / L_{ee}$ 
in the FI9602 scan {\protect \cite{SND2}}.}
\label{LGGTOLEE}
\end{figure}

The luminosity is determined by the following expression for
each process:
$N_i = L_i \sigma _{oi}$ , where
$i$ is the fixed beam energy point number;
$N_i$- number of  $e^+e^- \to e^+e^-$ or
$e^+e^- \to \gamma \gamma$ events;
$\sigma _{oi}$- detection cross section, calculated with an accuracy up
 to the third order in $\alpha$.
The ratio of luminosities $L_{\gamma \gamma}/L_{ee}$ for the third
scan is shown in Fig.~\ref{LGGTOLEE}. In part of FI96 experiment
\cite{SND2} this ratio was biased from its ideal value by several percent
due to low gain in the inner drift chamber.


\section{ Physical results of FI96 experiment }

The FI96 experiment itself was described in our previous preprint
\cite{SND2}.
In total, six successive scans were performed ($FI9601 \div FI9606$)
with the total integrated luminosity of $\sim 4~pb^{-1}$, corresponding to
approximately 6.5 million $\phi$- mesons produced. At present, more
than a half of the total recorded data are available for analysis
(scans $FI9602 \div FI9604$).


\subsection{ General description of the SND data analysis }

The SND data processing procedure consists of several successive 
steps \cite{DAQ}.
\begin{enumerate}
\item Reconstruction of events, stored on primary tapes. For each event
a list of particles with their parameters, including energies, angles, etc.,
is built. Reconstructed events,
together with the parameter lists, are written to secondary tapes.

\item Scanning of secondary tapes and creation of third-generation tapes,
 containing certain event classes, e.g. events containing
6--7 photons, or events with two charged particles and three photons,
etc.
\item Events of the classes, mentioned above are downloaded onto hard disks
for further processing using well known application packages PAW \cite{PAW}
and MINUIT \cite{MINUIT}, together with packages, developed in BINP: COCHA
data management system \cite{COCHA}, GIST histogram
code \cite{GIST}, UNIMOD Monte Carlo simulation code \cite{UNIMOD}, 
ART tape archiving system \cite{ART} , and other programs.

\end{enumerate}

The GIST histogram package calculates parameters, specific for the SND.
Let us list some of them, the most widely used in data processing:

$E_{tot}/2E$ --- total energy deposition in the calorimeter, divided by the center
of mass energy $2E$;

$E_{np}/2E$ --- sum energy deposition in the calorimeter by neutral
particles, measured in the units of the center of mass energy $2E$;

$P_{tot}/2E$ --- absolute value of the sum of all particles momenta
in an event normalized by a center of mass energy;

$N_\gamma$ --- number of photons, found in an event;

$N_{cp}$ --- number of charged particles;

$\vartheta _i$ --- polar angle of the $i$-th particle
(particles are sorted in the following way:
charged particles first, then neutral particles, with
descending order in energy within each group);

$\vartheta _{min}$ --- minimal spatial angle between particles
and beam axis;

$\varphi _i$ --- azimuth angle of $i$-th particle;

$R_i$ --- distance between the i-th particle track and beam axis in
$R-\varphi$- plane;

$Z_i$ --- $Z$ - coordinate of the i-th particle track intersection with
beam axis in $R - Z$- plane;

$\chi ^2 _E$ --- parameter, characterizing quantitatively the degree to
which the energy-momentum conservation law is held in an event,
analogous to $\chi ^2$ in statistics;

$\chi ^2 _m$ --- parameter, characterizing the degree of likelihood of
assumption, that there are intermediate $\pi^o$ or $\eta$- mesons in
an event;

$\chi ^2 _{\gamma}$ --- parameter characterizing quality of photons in
an event \cite{BEK_IV}, \cite{BOZ_IV_SIL}.

The total number of parameters available under GIST package is more than
2000. The parameters from this set could be further combined to produce
complex constraints and logical formulas for event selection.

During study of processes with substantial statistics the distributions
of selected events over beam energy were fitted using the following expression:

\begin{equation}
N_i = L(E_i) \cdot ( \sigma (E_i) \cdot
\delta(E_i) \cdot
\delta_{beam}(E_i) \cdot
\varepsilon(E_i) + \sigma _B(E_i) ),
\label{CROSSSECTION}
\end{equation}

where  $N_i$ is a  number of events in the $i$-th energy point
$L$ --- integrated luminosity in this energy point;
$\sigma $ --- theoretical cross section of the process under study;
$\delta$ ---  factor, taking into account radiative corrections, calculated
for each energy point by convolution of theoretical energy dependence of
the cross section with the probability for a photon with a certain
energy to be emitted
\cite{RADCORR}; $\delta _{beam}$ --- factor, accounting for beam energy
spread, also obtained by convolution of radiative corrected cross section
with a Gaussian energy distribution of the beam particles;
$\varepsilon$ --- detection efficiency for a process under study for
actual event selection criteria, calculated using Monte Carlo simulation;
$\sigma _B$ --- total visible cross section for all background processes,
calculated in the same way as for the process under study.

Also were taken into account beam energy corrections in individual
scans.

The dependence of theoretical cross section on the beam energy for
the processes proceeding via intermediate vector resonances
$V=\rho$, $\omega$, $\phi$ was approximated as follows
(see also \cite{ND}):

\begin{equation}
\sigma (E) = \arrowvert \sum_{i=\rho,\omega,\phi} \sqrt{\sigma_{0 V_{i}}}
\cdot \frac{m_{V_i} \Gamma_{V_i} e^{i \delta_{V_i}}}{D_{V_i}(S)} \arrowvert ^2
\label{CROSSECTION}
\end{equation}.

Here $\sigma _{oV}$ is a peak cross section;
$m_V$, $\Gamma _V(s)$ are mass and width of the resonance respectively,

$D_V(s) =m_V^2-s-i \sqrt{s} \Gamma _V(s)$;
$s=4E^2$.

The formulae describing the energy dependence of the resonance
widths ($\Gamma _V(s)$) and corresponding references are cited in
\cite{ND}.


%

To describe the interference between resonant and nonresonant processes
the following formula was used:

\begin{equation}
\label{CS_INTERF}
\sigma (E) = \sigma _o(1+\sigma '(2E-m_{\phi})) \cdot
\arrowvert 1-Z \frac{m_{\phi} \Gamma _{\phi}}{D_{\phi}} \arrowvert ^2,
\end{equation}

where $\sigma '$ is a derivative of the nonresonant cross section over energy;
$Z$ is a complex amplitude of the resonant process.


\subsection{ Measurements of $\phi$- meson parameters
in the process $\phi \to K_S K_L \to $ $neutral$ $particles$ }

The main goal of the study was to obtain the values of
 $m_{\phi}$, $\Gamma _{\phi}$, $\sigma _o$, and
$B(\phi \to K_S K_L)$
with lowest possible statistical error in order to use them in the analyses of
other processes with smaller statistics and for evaluation of
systematic errors. Given the integrated luminosity of $3.4~pb^{-1}$
and detection efficiency of $\varepsilon \sim 15\%$, the number of
detected $\phi \to K_SK_L \to neutral$ $particles$ events 
could be estimated to be equal to $\sim 10^5$, which leads to a statistical
error in a resonance mass: 
$\sim \Gamma _{\phi }/2\sqrt{N_{\phi }} \sim 0.01 MeV$.
Since the FI96 experiment consisted of six successive scans of the
$\phi$- region, the study of the $K_SK_L$ production makes
possible to check the energy calibration of the collider. Usually
the energy scale of the collider depends on many factors. To a 
relatively good accuracy of better than $20~keV$ the beam energy can be
expressed in the following way: 
$E(B,T)=k\cdot B + r\cdot T + C$, where 
$B$ and $T$ are the magnetic field strength and the collider temperature,
$k$, $r$, $C$ are the parameters to be determined during collider
beam energy calibration. The correction accounting for  variations
of the collider beam
circulation frequency $\Delta E/E = - 6\cdot \Delta f/f$
was also included.

\begin{figure}[htb]
  \begin{center}
    \mbox{\epsfig{figure=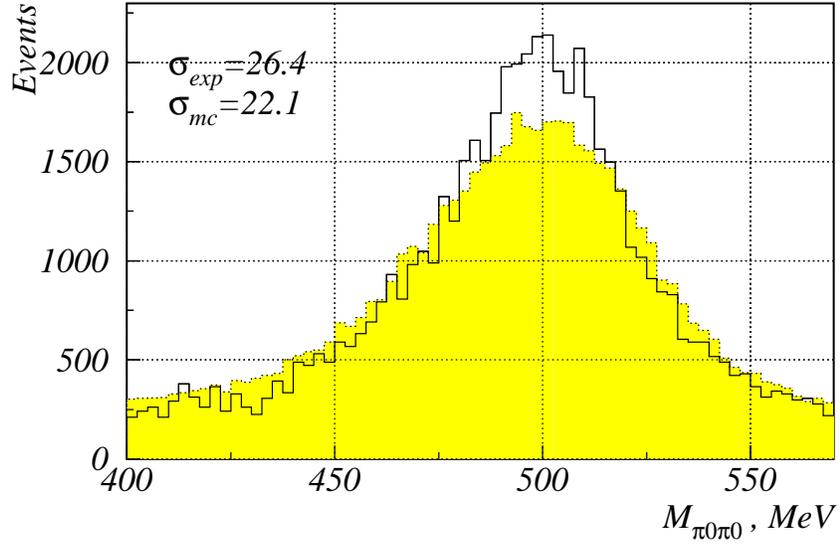,%
                  width=\textwidth}}
  \end{center} 
\caption{Invariant mass distribution of $\pi^o$- meson pairs from the
$K_S\to 2\pi^o$ decays.}
\label{DIMA_MKS}
\end{figure}

\begin{figure}[htb]
  \begin{center}
    \mbox{\epsfig{figure=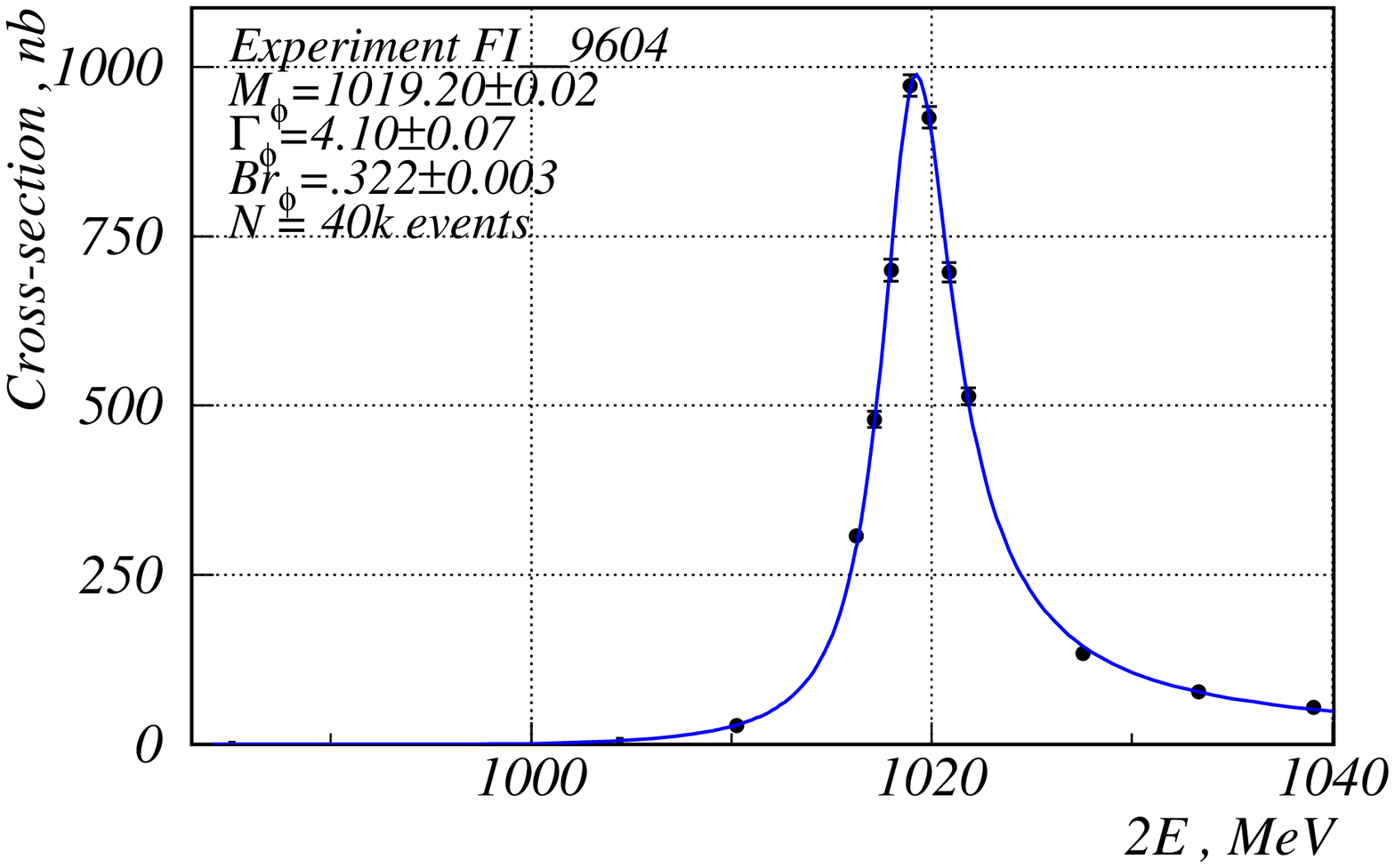,%
                  width=\textwidth}}
  \end{center} 
\caption{Cross section of the $e^+e^- \to \varphi \to K_SK_L$
process.}
\label{DIMA_CS_KSKL}
\end{figure}

Events were selected with $N_\gamma \geq 4$, $0.4<E_{tot}/2E<1.2$,
with detected
$K_S \to 2\pi^o$ decay, and measured 
$m_{K_S}$ within $400 \div 600 MeV$ interval (Fig.~\ref{DIMA_MKS}).
Detection efficiency for the $\phi \to K_SK_L$ decay is equal to
$\varepsilon = 14.8\%$. For the background  $\phi \to \eta \gamma$ process,
imitating the process under study due to misidentification of photons, it is
equal to $\varepsilon _1= 1.1\%$. Experimental data were approximated as a
sum of the process under study, resonant background from
$\phi \to \eta \gamma$, and nonresonant background.
The integrated luminosity was measured using 
$e^+e^- \to \gamma \gamma$ process. The resonance excitation curve
is shown in Fig.~\ref{DIMA_CS_KSKL} and the results of processing of three
scans are listed in the Table~\ref{DIMA_TAB}.

{\setlength{\tabcolsep}{3pt}
\begin{table}[!htb]
\caption{ $\phi$- meson parameters from the
$e^+e^- \to \phi \to K_SK_L$ process.}
\label{DIMA_TAB}

\begin{center}
\footnotesize

\begin{tabular}{||c|c|c|c|c||}

\hline
              & FI9602 & FI9603 & FI9604 & PDG,1996 \\
\hline
$m_{\phi}, MeV $
        & $1019.48\pm 0.03 $ & $1019.30\pm 0.03$ &$1019.20\pm 0.03$
        &$1019.413\pm 0.008$\\
$\Gamma_{\phi}, MeV $
        & $4.02\pm 0.10$ & $4.59\pm 0.08$ & $4.10\pm 0.08$ & $4.43\pm 0.05$\\
$\sigma_o, nb $ & $1314\pm 17$ & $1322\pm 13$ & $1405\pm 14$ & $1488\pm 22$ \\
$B(\phi \to K_SK_L), \%$
& $30.1\pm 0.4$ & $30.3\pm 0.3$ & $32.2\pm 0.4$ & $34.1\pm 0.5$ \\
Number of events& 51247 & 78335 & 49871 & --- \\
$\chi ^2 / N_D$ & 10/11 & 43/12 & 12/13 & --- \\
\hline
\end{tabular}
\end{center}
\end{table}
}
Main conclusions from the Table~\ref{DIMA_TAB}:

\begin{enumerate}
\item There exists a discrepancy between $\phi$- meson mass estimations in
the second and fourth scans:
$\Delta m = 0.28 \pm 0.04 MeV$, which cause is yet unclear.

\item In both 2-nd and 4-th scans $\chi^2/N_D$ parameter is small,
which is an argument in favor of a good beam energy stability during these 
scans, at least in the vicinity of the $\phi$- meson peak.

\item A beam energy leap presumably occurred in the FI9603 scan. This may
explain larger $\phi$- meson width and high $\chi^2/N_D$ ratio in this scan.

\item Taking into account the former, only FI9602 and FI9604 scans were
used for $\phi$- meson width determination. The combined result is
$\Gamma _\phi = 4.06 \pm 0.08 MeV$. This is by 3 standard deviations
smaller than its table value. The corresponding values of decay parameters
are the following:

$\sigma _o =(1360 \pm 11 \pm 45) nb$

$B(\phi \to K_SK_L) = (31.0 \pm 0.3 \pm 1.0 )\%$.
\end{enumerate}

We hope, that processing of the next three scans will clarify situation
with systematic errors due to collider energy stability and detector
performance.


\subsection{ Radiative decays }
\subsubsection{ The $\phi \to \eta \gamma$ decay}

The $\phi \to \eta \gamma$ decay is a classic magnetic dipole transition
from  $\phi$- into $\eta$- meson. In terms of nonrelativistic quark
model  \cite{DONNELL} this decay can be described as a quark spin flip
inside  $\phi$- meson: $^3S_1 \to ^1S_0 + \gamma$. The theoretical
estimation and experimental value agree within about $10 \%$.
Up to now, more than 10 measurements of $\phi \to \eta \gamma$ branching
ratio were done and its current table value is equal to $1.26 \pm 0.06\%$.
In this paper preliminary results of new measurements of this
parameter in three $\eta$- meson decay modes are presented:
$\eta \to 2\gamma$ ($B(\eta \to 2\gamma)=39\%$);
$\eta \to \pi^+\pi^-\pi^o$ ($B(\eta \to \pi^+\pi^-\pi^o)=23\%$) ט
$\eta \to 3\pi^o$ ($B(\eta \to 3\pi^o)=32\%$).

{\center \bf The $\phi \to \eta \gamma \to 3\gamma$ channel}

\begin{figure}[htb]
  \begin{center}
    \mbox{\epsfig{figure=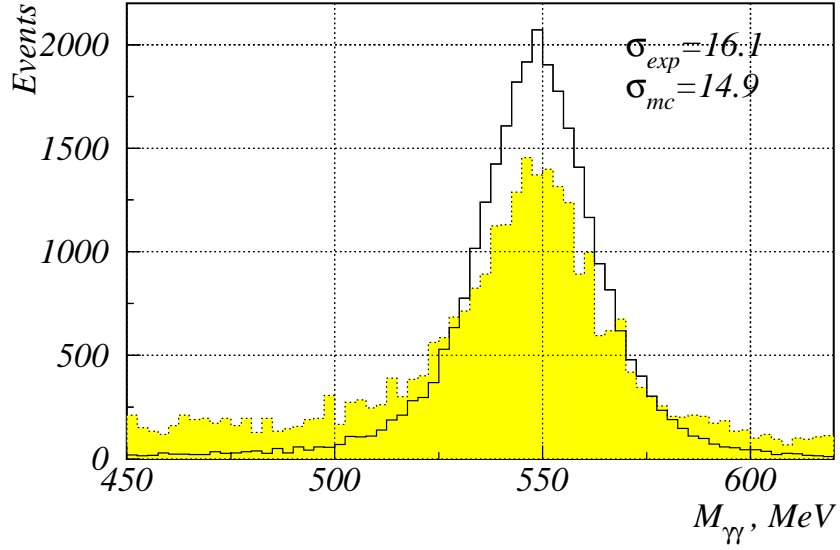,%
                  width=\textwidth}}
  \end{center} 
\caption{$\eta$- meson invariant mass in the
$e^+e^- \to \varphi \to \eta \gamma$ process.}
\label{DIMA_META}
\end{figure}

The following event selection criteria were used:
$N_{\gamma}=3,4$; $0.8<E_{tot}/2E<1.1$; $P_{\perp}/2E<0.175$;
$\vartheta _{min} > 24^o$; $E_{\gamma min}>0.2E$;
$550<m_{\gamma \gamma}<800 MeV$; $\chi^2_E<30$.
Events from two $\phi$- meson scans were processed.
Simulated events of the 
$\phi \to \eta \gamma$; $\phi \to K_SK_L$; $\phi \to \pi^o \gamma$;
were processed in the same way in order to evaluate their
detection efficiencies, which turned out to be:
22\%, 0.01\%, and 8\% respectively.
In Fig.~\ref{DIMA_META} the $\eta$- meson invariant mass distribution 
is shown.

\begin{figure}[htb]
  \begin{center}
    \mbox{\epsfig{figure=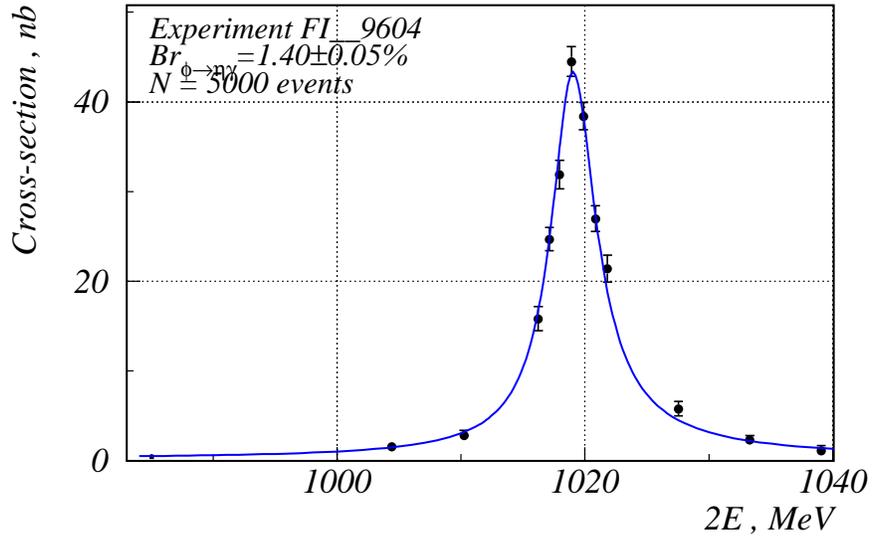,%
                  width=\textwidth}}
  \end{center} 
\caption{Cross section of the $e^+e^- \to \varphi \to \eta \gamma$
process.}
\label{DIMA_CS_ETAG}
\end{figure}

The cross section energy dependence was approximated according to
formulae from \ref{CROSSSECTION}, \ref{CROSSECTION} with a resonant
background from the $\phi \to K_SK_L,\pi^o \gamma$ processes and a
nonresonant background taken into account. The
$\phi$- resonance  excitation curve in the $\phi \to \eta \gamma$
channel is shown in Fig.~\ref{DIMA_CS_ETAG}. The values of fit parameters
are listed in Table~\ref{DIMA_TAB_ETAG}.

\begin{table}[!htb]
\caption{ Main $\phi$- meson parameters measured using the
$e^+e^- \to \phi \to \eta \gamma \to 3 \gamma$ process.}
\label{DIMA_TAB_ETAG}

\begin{center}
\footnotesize

\begin{tabular}{||c|c|c|c|c||}

\hline
Experiment  & FI9602  & FI9604 & PDG,1996 \\
\hline
$m_{\phi}, MeV $ & 1019.48  & 1019.2  & $1019.413\pm 0.008$\\
$\Gamma_{\phi}, MeV $ & 4.02   & 4.1  & $4.43\pm 0.05$\\
$\sigma_o(e^+e^- \to\phi \to \eta \gamma), nb $ & 58.9$\pm$1.9
                                                & 59.6$\pm$2.2
                                                & 53.2$\pm$2.5 \\
$B(\phi \to \eta \gamma), \%$ & 1.34$\pm$0.05
                              & 1.40$\pm$0.05 & $1.26\pm 0.06$ \\
Number of events   & 4256  & 4040  & --- \\
$\chi ^2 / N_D$ & 16/10 & 20/12 & --- \\
\hline
\end{tabular}
\end{center}
\end{table}
Combining data from both scans one can obtain the following branching
ratio:

$B(\phi \to \eta \gamma) = (1.37 \pm 0.04 \pm 0.08) \%$,

where the first error is a statistical one and the second is
systematic.


{\center \bf The $\phi \to \eta \gamma \to 3\pi^o \gamma $ channel}

\begin{figure}[htb]
  \begin{center}
    \mbox{\epsfig{figure=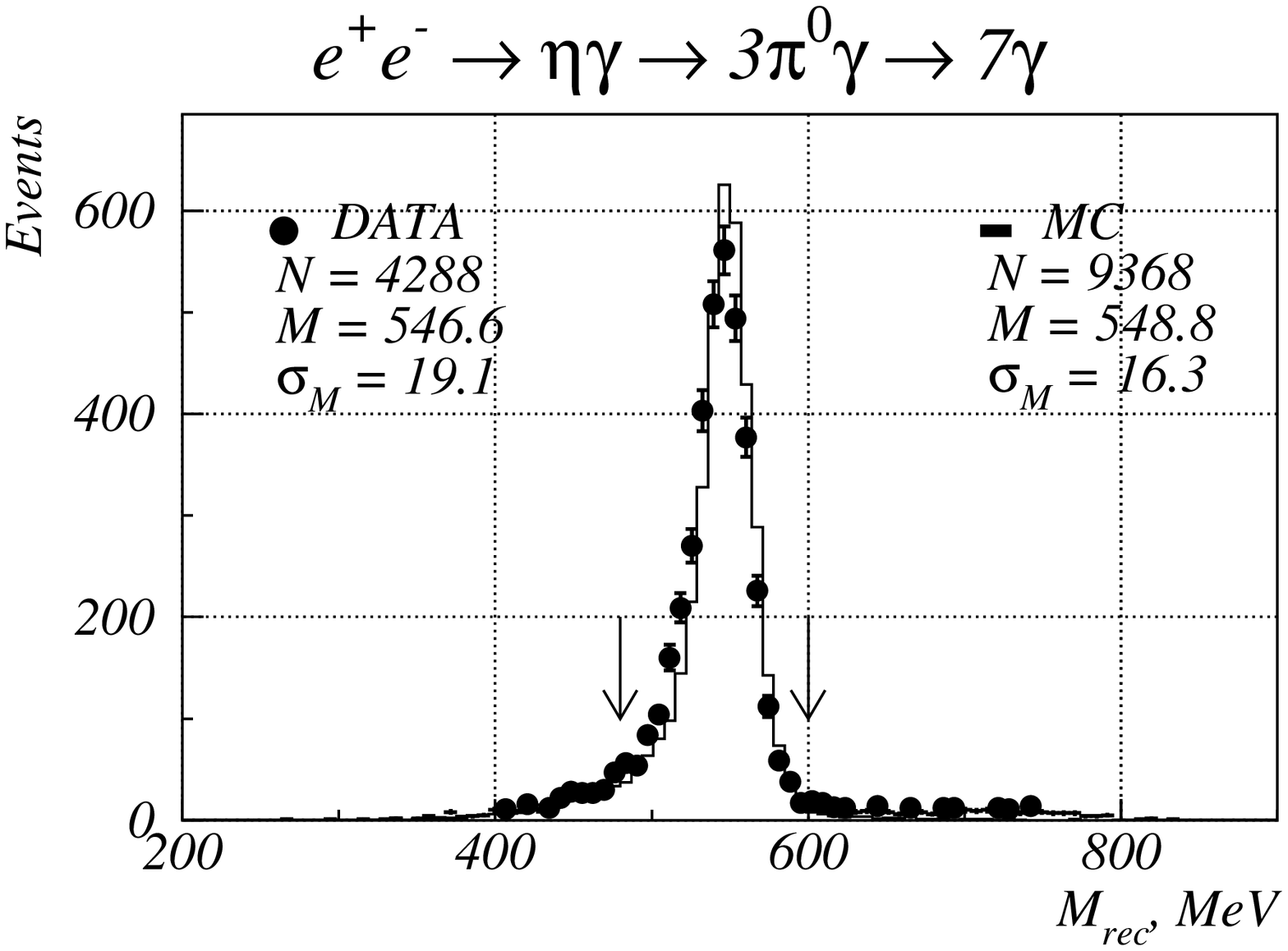,%
                  width=\textwidth}}
  \end{center} 
\caption{Recoil mass distribution for the most energetic
photon.}
\label{META_IV}

  \begin{center}
    \mbox{\epsfig{figure=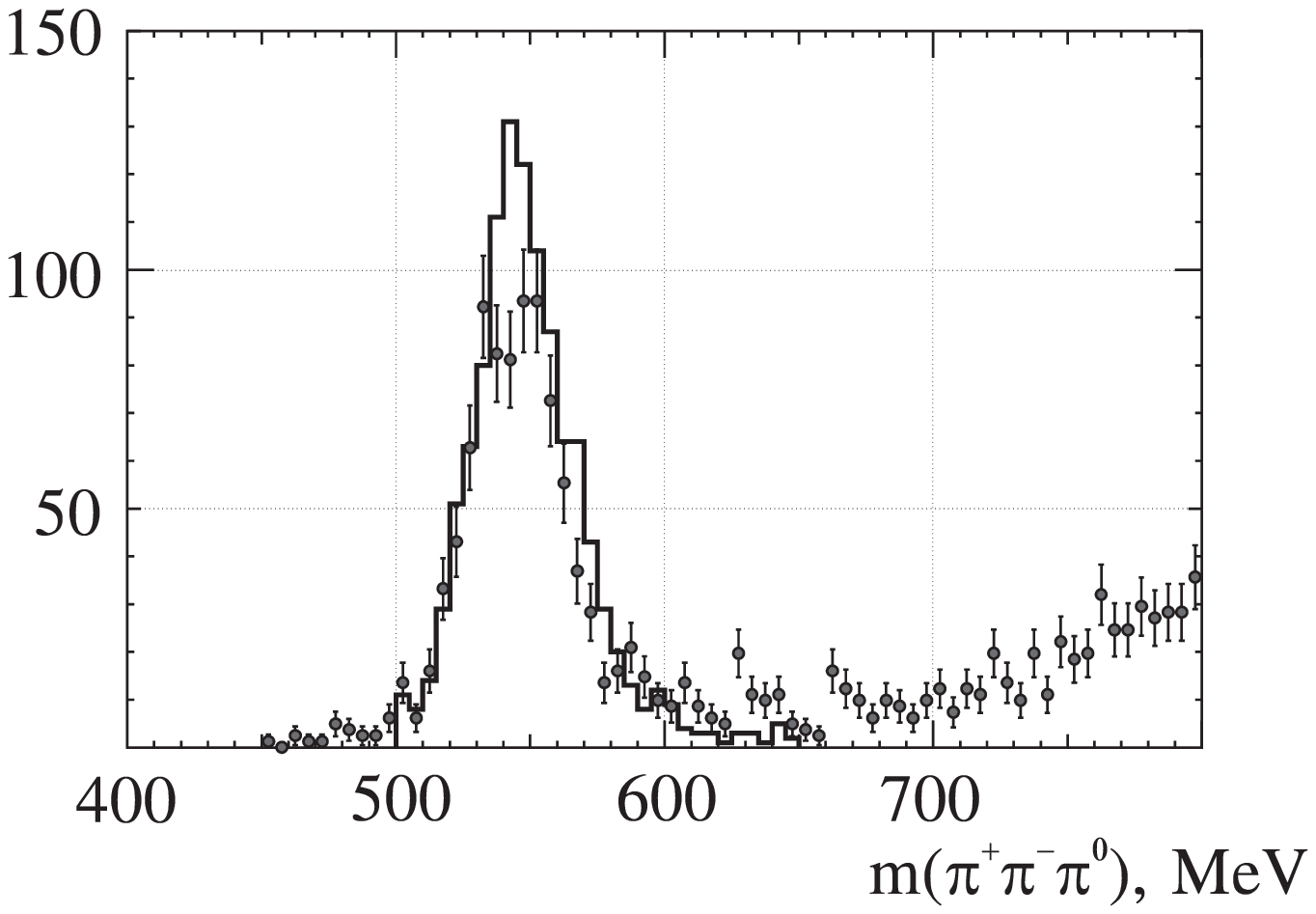,%
                  width=\textwidth}}
  \end{center} 
\caption{Distribution over the reconstructed mass of the $\eta$- meson;
points with error bars --- experiment, $\sigma _{exp}=16.6~MeV$;
histogram --- simulation, $\sigma _{sim}=16.9~MeV$.}
\label{META_BUK}
\end{figure}

Due to high efficiency of the SND calorimeter to multiphoton events, full
reconstruction of $\phi \to \eta \gamma \to 3\pi^o\gamma \to 6,7\gamma$
turned out to be feasible. The characteristic feature of this channel is a
peak at $\eta$- meson mass  in the recoil spectrum of the most energetic
photon in an event (Fig.~\ref{META_IV}).
Events were selected according to following criteria:
$N_\gamma =6,7$; $E_{tot}/2E=0.8\div 1.2$; $P_{tot}/2E<0.12$;
$\vartheta _{min}>27^o$, $\chi^2_E<25$, $\chi^2_{\gamma}<20$.
For the energy dependence of the cross section the standard approximation was used
(\ref{CROSSECTION}). The energy dependence of the detection efficiency was
neglected.
The background of $\sim 2~\%$ due to the process $\phi \to K_S K_L$ was 
subtracted using 
the events outside of the main peak in the Fig.~\ref{META_IV}.

In order to estimate SND systematic errors for multiphoton events
with full reconstruction of all photons,
an independent analysis of events with  $N_\gamma =7$ was carried out.
In these events the presence of three $\pi^o$- mesons was required.
The detection efficiency in this case was lower (Table~\ref{TAB_VI}),
but no background from other $\phi$- meson decays was observed at 
a present level of experimental statistics.

\begin{table}[!htb]
\caption{Results of the study of
$\phi \to \eta \gamma \to \pi^o \pi^o \pi^o \gamma$ decay
for three independent scans of  the $\phi$- resonance region.}
\label{TAB_VI}

\begin{center}
\footnotesize

\begin{tabular}{||c|c|c|c||}

\hline
              & FI9602 & FI9603 & FI9604 \\
\hline
Number of events $6,7\gamma$ & 1167 & 1748 & 1036 \\
$N_{7\gamma}/N_{6\gamma}$ (experiment)
                & $0.60\pm 0.04$ & $0.62\pm 0.04$ & $0.62\pm 0.05$ \\
$N_{7\gamma}/N_{6\gamma}$ (simulation)
                & $0.60\pm 0.02$ & $0.60\pm 0.02$ & $0.59\pm 0.02$ \\
Number of fully $7\gamma$ &  360 &  564 & 315 \\
reconstructed events&      &      &      \\
Efficiency $6,7\gamma $& $8.0\pm 0.1$   & $7.9\pm 0.1$   &  $7.7\pm 0.1$ \\
Efficiency $7\gamma $   & $2.72\pm 0.07$ & $2.64\pm 0.07$ & $2.57\pm 0.07$\\
$B(\phi \to \eta \gamma )$ $6,7\gamma$, \%
                & $1.24\pm 0.04$   & $1.23\pm 0.03$   &  $1.31\pm 0.04$ \\
$B(\phi \to \eta \gamma )$ $7\gamma $, \%
                & $1.14\pm 0.06$   & $1.17\pm 0.05$   &  $1.20\pm 0.07$ \\
\hline
\end{tabular}
\end{center}
\end{table}

It is clearly seen from the Table~\ref{TAB_VI} that the cross sections
and branching ratios of the process
$\phi \to \eta \gamma \to 3\pi^o \gamma$ for the events with $N_\gamma =6,7$
and $N_\gamma =7$ differ by approximately 8\% ($2\sigma$).
This systematic error can be explained by misidentification of photons in
7 gamma events. 

The data for $N_\gamma$ =6,7 from all three scans were combined
and the following final result was obtained:

$\sigma _o(\phi \to \eta \gamma \to 3\pi^o \gamma) = 
(54.6 \pm 1.3 \pm 3.5) nb$

$B(\phi \to \eta \gamma) = (1.25 \pm 0.03 \pm 0.08) \% $.


{\center \bf The $\phi \to \eta \gamma \to \pi^+ \pi^- \pi^o \gamma$
channel}

Since this decay channel in addition to three photons contains two
charged particles in its final state, the
performance of the drift chambers can contribute into systematic
error of the result. Comparison of this result with those for pure neutral
channels can be a good probe of the tracking system performance.

Events were selected according to following criteria:
$N_{cp}=2$, $N_\gamma \geq 3$,
spatial angle between charged particles $\alpha _{1,2} < 150 ^o$.
Events must satisfy energy and momentum conservation laws with additional
requirement, that two photon pairs must have an invariant masses close
to that of $\pi^0$. This requirement was imposed within a kinematic fitting
procedure \cite{KINEMFIT}.
This procedure calculated the following parameters: the minimum of logarithmic
likelihood function $L_{\eta \gamma}$, $m_{\eta}$ --- the $\eta$- meson mass
estimate,
$D_L$ --- difference between minimums of logarithmic likelihood functions for 
two best combinations of photons, producing required  $\pi^0$-s,
$D_{L3\pi}$ --- difference in likelihood functions minimums in $\eta
\gamma$ and $\pi^+ \pi^- \pi^o$ hypothesis.
In Fig.~\ref{META_BUK} experimental and simulated distributions over
reconstructed mass of the $\eta$-meson are shown. 
After addition of two more requirements:
$L_{\eta \gamma}<15$,
$500 < m_{\eta} < 600 MeV$,
the detection efficiency was estimated to be  $30.4\pm0.9\%$, or,
taking into account  $B(\eta \to \pi^+ \pi^- \pi^o)=0.23$, was equal to
$\varepsilon (\eta \gamma \to \pi^+ \pi^- \pi^o \gamma)=(7.0\pm 0.2)\%$

Processing of the FI9602 scan gave the following results on peak
cross section and decay branching ratio:

$\sigma _o(e^+e^- \to \phi \to \eta \gamma ) =
42.7 \pm 3.1 \pm 4.4 nb$

$B(\phi \to \eta \gamma) = (0.98 \pm 0.07 \pm 0.10) \%$.

Only background subtraction error and
statistical error of simulation were included into the systematic error.

The data were also processed under more stringent selection criteria:
$N_{cp}=2$, $N_\gamma \geq 3$,
$\alpha _{1,2} < 130 ^o$,
$R _{1,2} < 0.5~סל$,
$L_{\eta \gamma}<10$,
$D_L >2$,
$D_{L3\pi } < -0.5$,
$500 < m_{\eta} < 600~MeV$.
The detection efficiency here is smaller,
$\varepsilon (\eta \gamma \to \pi^+ \pi^- \pi^o \gamma)=5.4\pm 0.2\%$,
but the background was additionally reduced by a factor of 3.
Thus, the background subtraction error here was significantly lower.

$\sigma _o(e^+e^- \to \phi \to \eta \gamma ) =
(48.8 \pm 3.8 \pm 3.8) nb$

$B(\phi \to \eta \gamma) = (1.04 \pm 0.08 \pm 0.08) \%$.

Since the systematic error of the latter result is estimated to be
smaller, we consider the last result as a conclusive one for this channel.

{\center \bf Conclusions on the $\phi \to \eta \gamma$ channel}

Results obtained for these three different channels do not agree well.
Additional study of systematic errors is necessary. Taking into account
these differences and common systematic errors,
the following combined results can be obtained:

$\sigma _o(e^+e^- \to \phi \to \eta \gamma ) =
(56.8 \pm 2.6\pm 3.1) nb$

$B(\phi \to \eta \gamma) = (1.30 \pm 0.06 \pm 0.07) \%$.

\subsubsection{ The $\phi \to \pi^o \gamma$ decay}

The $\phi \to \eta \gamma$ decay was studied on the basis of one scan
of $\phi$ -meson region, corresponding to a total integrated luminosity
of $0.4 pb^{-1}$ i.e.$10\%$ of experimental statistics. The preliminary
result was already published in \cite{SND2} and is equal to

$B(\phi \to \pi^o \gamma) = (0.10 \pm 0.02)\% $.


\subsection{ Rare radiative decays }
\subsubsection{ Search for the  $\phi \to \eta ' \gamma $ decay }

Radiative decay $\phi \to \eta ' \gamma$ is a good probe of an internal
structure of the $\eta '$- meson \cite{ROSNER,GILMAN,DONNELL}.

As it was mentioned in \cite{ROSNER}, the  $\phi \to \eta ' \gamma$
decay branching ratio strongly depends on the gluonium  and strange quarks
contents in $\eta '$- meson. In an assumption of an
ordinary quark structure of  $\eta '$,
predicted branching ratio is $B(\phi \to \eta ' \gamma) = 7\cdot 10^{-5}$.

Two attempts to measure this decay were done earlier. The first upper
limit was obtained with Neutral Detector \cite{ND}:
$B(\phi \to \eta ' \gamma) < 4.1\cdot 10^{-4}$; and recently was
published a preliminary result from CMD-2 detector \cite{CMD2_PROCEED}:
$B(\phi \to \eta ' \gamma) < 2.4\cdot 10^{-4}$.

In this work the process
\begin{equation}
\phi \to \eta ' \gamma \to \pi^+ \pi^- \eta \gamma \to
\pi^+ \pi^- \pi^+ \pi^- \pi^o \gamma \to \pi^+ \pi^- \pi^+ \pi^- 3 \gamma
\label{proc_2}
\end{equation}
was studied.

Among all decay modes of $\eta '$- meson this one has a relatively
high probability:
$B(\eta ' \to \pi^+ \pi^- \eta) \cdot
B(\eta  \to \pi^+ \pi^- \pi^o) \sim 10\%$,
well defined final state, practically no combinatorial background in
neutral particles, and it also contains an $\eta$ meson in the decay chain.


Event selection for the search of the process (\ref{proc_2}) was
done using the following basic set of conditions:
$N_{cp}=4$; $N_{\gamma}=3$; $R_{cp}<0.3~cm$; $|Z_{cp}|<6~cm$;
number of hit wires in the drift chambers corresponds to four
tracks. For the events surviving those cuts kinematic fit was performed
with additional requirements: $0.1<E_{np}/2E<0.3$; energy and momentum
conservation is held to the accuracy higher than $10~MeV$; minimal spatial angle
between any two particles $\alpha _{min}$ is greater than $9$ degrees;
total energy deposition of charged particles in the second calorimeter
layer $\Delta E_{i2}$ is less than $40~MeV$
 
In addition, events were required to have some fixed kinematic
parameters due to presence of intermediate particles in the decay:
recoil mass of one of the photons must be equal to $\eta'$ mass;
invariant mass of the other two photons -- equal to $\pi^o$ mass;
and invariant mass of this $\pi^o$ together with two charged pions --
equal to $\eta$- meson mass. These masses were calculated for analysis,
but no additional cuts were imposed,

Detection efficiency under these selection criteria was determined from
simulation to be equal to $\varepsilon = (0.57 \pm 0.07)\%$.

Currently the data from three scans $FI9602\div FI9604$ were processed,
corresponding to a total number of produced $\phi$- mesons
$N_{\phi} \sim 4.1\cdot 10^6$.

\begin{figure}[htb]
  \begin{center}
    \mbox{\epsfig{figure=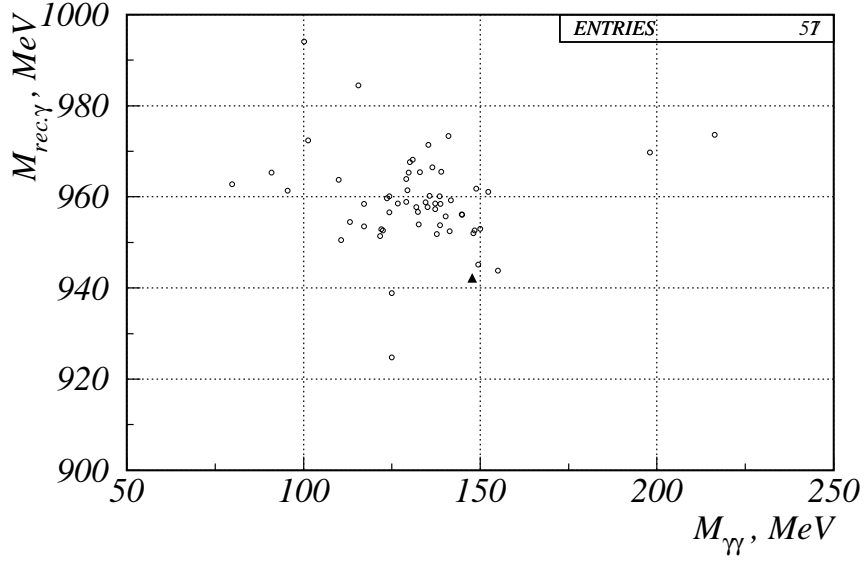,%
                  width=\textwidth}}
  \end{center}
\caption{$m_{rec.\gamma}$ vs. $m_{\gamma \gamma}$ scatter plot
for the $e^+e^- \to \varphi \to \eta ' \gamma \to 
\pi^+\pi^-\pi^+\pi^-3\gamma$ decays. Circles --- simulation,
triangle --- the event, satisfying selection criteria.}
\label{MYPLOT}
\end{figure}

Only one event satisfying all intermediate states
requirements was found (Fig.~\ref{MYPLOT}). Corresponding branching ratio is
equal  $B(\phi \to \eta ' \gamma) = 4\cdot 10^{-5}$.

The statistics of simulated events
 of main background processes, available by now, is too 
low to estimate their contributions and only upper limit corresponding to one 
detected event could be placed:
$B(\phi \to \eta ' \gamma) < 1.7\cdot 10^{-4}$
at 90\% confidence level. 


\subsubsection{ Study of electric dipole decays of light vector mesons}

Electric dipole transitions are widespread in atoms, but in mesons,
composed of light quarks transitions of the type $V \to S \gamma$
$(^3S_1 \to ^3P_o + \gamma )$, where
 $V$ is a vector meson
($\rho$, $\omega$, $\phi$) and $S$ is a scalar one ($a_o$, $f_o$, $\ldots$) 
are strongly
suppressed by a phase space factor because even the lightest scalar mesons
are quite heavy ($M \sim 1~GeV$). 

Quark structure of scalar mesons is not conclusively determined yet.
In a standard two-quark description $a_o$ ט $f_o$ mesons look like
scalar analogs of $\rho$- and $\omega$- mesons:

$a_o=\frac{u\bar{u}-d\bar{d}}{\sqrt{2}}$ and
$f_o=\frac{u\bar{u}+d\bar{d}}{\sqrt{2}}$ or $s\bar{s}$.

Better chances has now a four-quark model:

$a_o=\frac{u\bar{u}-d\bar{d}}{\sqrt{2}}s\bar{s}$ and
$f_o=\frac{u\bar{u}+d\bar{d}}{\sqrt{2}}s\bar{s}$  \cite{U1}.

Yet another model was suggested -- kaon molecule, according to which,
$a_o$ ט $f_o$ are bound states of two $K$- mesons (\cite{CLOSE}, \cite{U2}).

And finally, real states can include both ordinary and exotic contributions.
The probability of transition $\phi \to S \gamma$ strongly depends on
quark structure of mesons involved in the decay. Thus the idea emerged
\cite{U3} to use radiative decays as a probe in order to clarify the
structure of  $a_o$- ט $f_o$- mesons.
The results of estimations of branching ratios in different models and
the experimental values are listed in Table~\ref{TAB_SCAL}.
The most important result is that four-quark model predicts branching
ratios almost an order of magnitude higher than two-quark and kaon molecule
models.
{\setlength{\tabcolsep}{2pt}
\begin{table}[!htb]
\caption{Expected values of scalar meson decays branching ratios
from the work  {\protect \cite{TAB_IV_PAK,U3}} 
and experimental results.}
\label{TAB_SCAL}

\begin{center}
\footnotesize

\begin{tabular}{||c|c|c|c||c|c|c||}

\hline
       & $q\bar{q}$ $(s\bar{s})$  &  $q\bar{q}q\bar{q}$ & $K\bar{K}$
       & ִֽ \cite{ND} & CMD-2 \cite{CMD2_PROCEED} & This work \\
\hline
$B(\phi \to f_o \gamma \to \pi^o \pi^o \gamma)$
   & $5\cdot 10^{-5}$ &  $10^{-4}$  &   $10^{-5}$
   & $<2\cdot 10^{-3}$ & $<7\cdot 10^{-4}$ & $(4.7\pm 1.0)\cdot 10^{-4}$\\
$B(\phi \to a_o \gamma \to \eta  \pi^o \gamma)$
   & $8\cdot 10^{-6}$ &  $10^{-4}$  &   $10^{-5}$
   & $<2.5\cdot 10^{-3}$ & --- & $(1.3\pm 0.5)\cdot 10^{-4}$  \\
\hline
\end{tabular}
\end{center}
\end{table}
}

The first time a search for electric dipole transitions was
carried out with ND detector \cite{ND},
where upper limits at a level of $\sim 10^{-3}$ were established.
At present the search for the decays of the kind of
$\phi \to S \gamma$ is being performed at VEPP-2M collider by
two detectors: SND and CMD-2. A new $\phi$- factory is close
to commissioning in Frascati \cite{DAFNE}, where the search for
such decays was stated as one of the main goals. Another onslaught
on neutral decays of $\phi$- meson is being prepared at a CEBAF
photon beam \cite{CEBAF}. So, during next few years one can expect,
that situation with these decays will be greatly clarified.


{\bf \center Evidence of the $\phi \to \pi^o \pi^o \gamma$ decay}

When studying the

\begin{equation}
\label{IVB1}
e^+e^- \to \phi \to \pi^o \pi^o \gamma
\end{equation}

channel, one should be aware about significant background contributions,
coming from the following processes:

\begin{equation}
\label{IVB2}
e^+e^- \to \phi \to \eta \gamma \to 3\pi^o \gamma,
\end{equation}

\begin{equation}
\label{IVB3}
e^+e^- \to \omega \pi^o \to \pi^o \pi^o \gamma,
\end{equation}

\begin{equation}
\label{IVB4}
e^+e^- \to \rho \pi^o \to \pi^o \pi^o \gamma,
\end{equation}

\begin{equation}
\label{IVB5}
e^+e^- \to \phi \to K_SK_L \to neutral  \  particles.
\end{equation}

\begin{figure}[htb]
  \begin{center}
    \mbox{\epsfig{figure=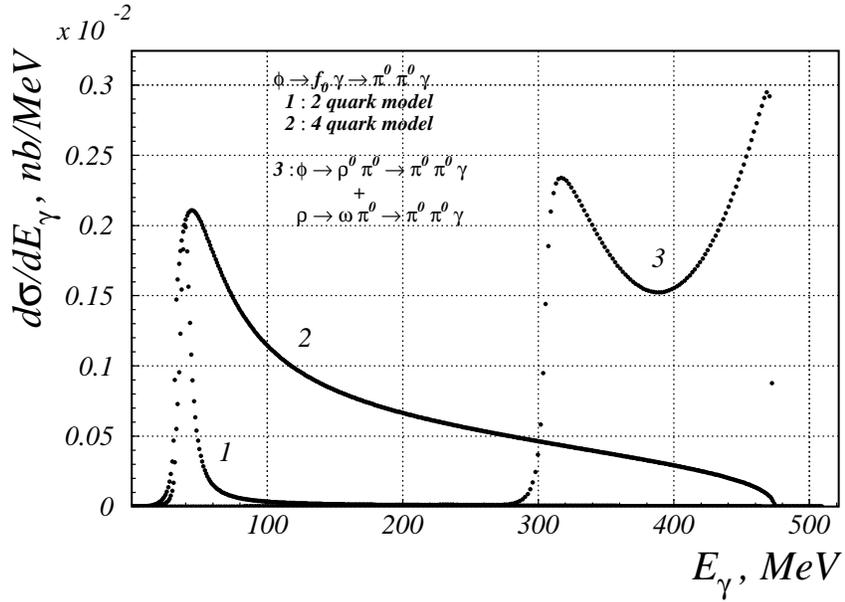,%
                  width=\textwidth}}
  \end{center}
\caption{Recoil photon spectrum in the process 
$e^+e^- \to \pi ^o \pi ^o \gamma $.}
\label{IVB_ERECG}
\end{figure}

Fig.~\ref{IVB_ERECG} shows expected recoil photons spectra in the
processes (\ref{IVB1}), (\ref{IVB3}), (\ref{IVB4}).
One can see, that for $E_{\gamma}<250~MeV$ contributions
from the processes
(\ref{IVB3}, \ref{IVB4}) are significantly suppressed. The processes
(\ref{IVB2}, \ref{IVB5}) contribute only due to event misidentification,
for example, when photons merge together in the calorimeter.
Events were selected satisfying following criteria:
$N_{\gamma}=5$, $0.8 < E_{tot}/2E < 1.2$, $P_{tot}/2E < 0.12$,
$\chi^2_E < 25$. It was also required, that among all combinations of
photons existed one, in which two photon pairs have invariant
masses within the interval $m_{\pi^o} \pm 30 MeV$. 

\begin{figure}[htb]
  \begin{center}
    \mbox{\epsfig{figure=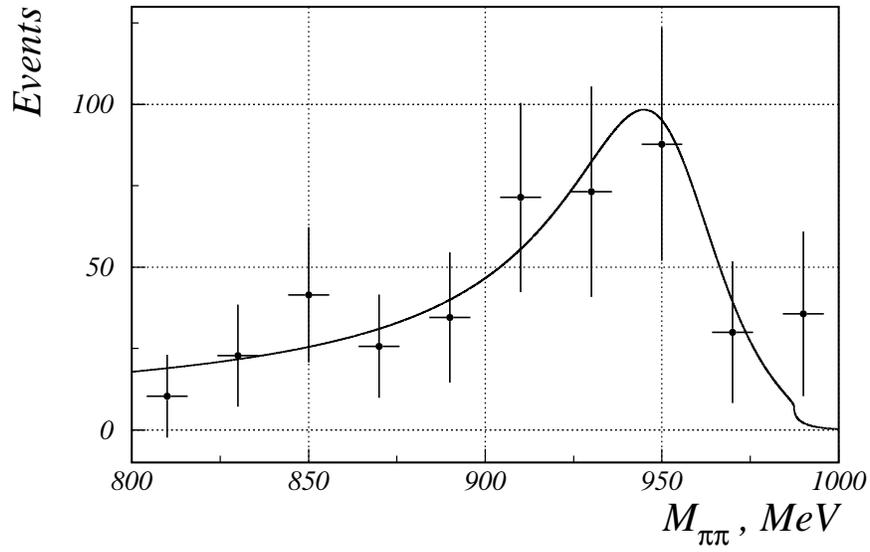,%
                  width=\textwidth}}
  \end{center}
\caption{Distribution over invariant mass of the
$\pi ^o \pi ^o$- system in the 
$e^+e^- \to \pi ^o \pi ^o \gamma $ process.}
\label{IVB_MP0P0}
\end{figure}

The processes (\ref{IVB2}, \ref{IVB5}) were suppressed with the help
of a special parameter, checking if the transverse profiles of
photons look like those of isolated ones ($\chi ^2_{\gamma} < 5$).
Then, the  events with a mass of $\pi^o \gamma$ -system lying within
the 720 --- 840~MeV interval were discarded in order to reject
contribution from the process (\ref{IVB3}). The $\pi^o \pi^o $ invariant
mass spectrum of the remaining events is shown in Fig.~\ref{IVB_MP0P0}.
Further analysis of the angular distribution revealed, that it
agrees with S- wave production mechanism of $\pi^o \pi^o $- system.
Detection efficiency for the process (\ref{IVB1}) under described selection
criteria decreases linearly with the increase of $m_{\pi^o \pi^o}$  from 14\%
at $m_{\pi^o \pi^o}=800 MeV$ to 6\% at $975 MeV$.
This dependence was obtained for the process (\ref{IVB1}) simulation with
two $\pi^o$- mesons in S- wave with radiative correction taken into account.
The number of events and expected background contributions are listed in
the Table~\ref{IVB_TAB}.

\begin{table}[!htb]
\caption{Number of the selected events
$e^+e^- \to \pi^o \pi^o \gamma$ with
$m_{\pi^o \pi^o} > 800 MeV$ and estimated background contributions. }
\label{IVB_TAB}

\footnotesize
\begin{center}

\begin{tabular}{||c|c||}

\hline
$N_{\phi}$ & $3.9\cdot 10^6$ \\
$\phi \to \pi^o \pi^o \gamma $, experiment  & 45 \\
$\phi \to \eta \gamma $, simulation  & 5 \\
$\phi \to K_SK_L $, simulation  & $< 6$ \\

$\phi \to \rho \pi^o, \omega \pi^o \to \pi^o \pi^o \gamma $
 , simulation & 1.4 \\
$B(\phi \to \pi^o \pi^o \gamma )$  & $(1.1 \pm 0.2)\cdot 10^{-4}$ \\
\hline
\end{tabular}
\end{center}
\end{table}

After background subtraction, in accordance with Table~\ref{IVB_TAB},
$38 \pm 7$ events remain, from which the following estimate for the
branching ratio could be derived:
$B(\phi \to \pi^o\pi^o\gamma ) = N_{\pi^o\pi^o\gamma}/
N_{\phi}\cdot \bar \varepsilon = (1.1 \pm 0.2) \cdot 10^{-4}$,
for $\pi^o \pi^o $- system mass above 800~MeV (only statistical error
indicated). So, in this experiment we observe the decay  (\ref{IVB1})
with S- wave production of $\pi^o$- pairs.
The comparison of $\pi^o \pi^o $ invariant mass spectra, depicted in 
Fig.~\ref{IVB_MP0P0}, with model estimations in Fig.~\ref{IVB_ERECG}
shows good agreement with a four-quark model predictions  \cite{U3}.

Fitting, using formulae of \cite{U3}, gives the following
$f_o$- meson parameters:

$m_{f_o} = (950 \pm 8) MeV$;

$g^2_{f_oKK}/4\pi = (2.3 \pm 0.5) GeV^{-2}$;

$g^2_{f_o\pi \pi}/4\pi = (0.4 \pm 0.1) GeV^{-2}$;

$B(\phi \to f_o(980) \gamma) = (4.7 \pm 1.0)\cdot 10^{-4}$.

The latter value was obtained 
within a framework of the 4-quark model (\cite{TAB_IV_PAK,U3}).
Also taken into account was the relation $B( f_o \to \pi^+ \pi^-) =
2 B( f_o \to \pi^o\pi^o)$. Systematic error in the branching ratio is
estimated to be about $ 20\%$. To improve it, one should account for the
background from the processes (\ref{IVB2}, \ref{IVB5}) more precisely
(look Table~\ref{IVB_TAB}).


{\bf \center Evidence of the $\phi \to \eta \pi^o \gamma $ decay }

        The main background for the process

\begin{equation}
\label{tenth}
e^+e^- \to \eta \pi^o \gamma
(\eta \to 2\gamma ,\pi^o \to 2\gamma )
 \to 5\gamma
\end{equation}

comes from the reactions (\ref{IVB2}), (\ref{IVB3}), (\ref{IVB5}).

The following event selection criteria were used:
$N_{\gamma} = 5$,
$\vartheta_{min} > 27^o$,
$E_{tot}> 0.8\cdot 2E_0$, $\chi^2_E < 20$, $\chi^2_M < 25$, $\chi^2_{\gamma}
< 20$, $NC25 < 6$. In the table \ref{tablepak1} the corresponding detection
efficiencies and number of events are listed.
Experimental statistics here is equal to $N_\phi = 6.5 \cdot 10^6$

\begin{table}[!htb]
\caption{ Detection efficiencies and expected number of events
for the main processes}
\label{tablepak1}

\begin{center}
\footnotesize

\begin{tabular}{||l|c|c||}

\hline  & Efficiency & Number of events \\

\hline Experiment &  --- & 283 \\

\hline $e^+ e^- \to \eta \pi \gamma$ & $\varepsilon = (8.7\pm0.3)\%$ &  --- \\  

\hline $e^+ e^- \to \eta \gamma$ & $(1.3 \pm 0.1) \cdot 10^{-3}$ & 
$109 \pm 9$ \\  

\hline $e^+ e^- \to K_S K_L$ & $ 5 \cdot 10^{-6}$ & $10 \pm 10$ \\ 

\hline $e^+ e^- \to \omega \pi$ & $0.4 \cdot 10^{-2}$ & $85\pm4$ \\

\hline 

\end{tabular}
\end{center}
\end{table}

After background subtraction we obtain the total number of events
$N(\phi \to \eta\pi^0\gamma)=79\pm 21$, which corresponds to a branching ratio:
$B(\eta\pi^0\gamma) = N/N_{\phi}\cdot \varepsilon = (1.40 \pm 0.35)\cdot 
10^{-4}$.

\begin{figure}[htb]
 \begin{center}
   \mbox{\epsfig{figure=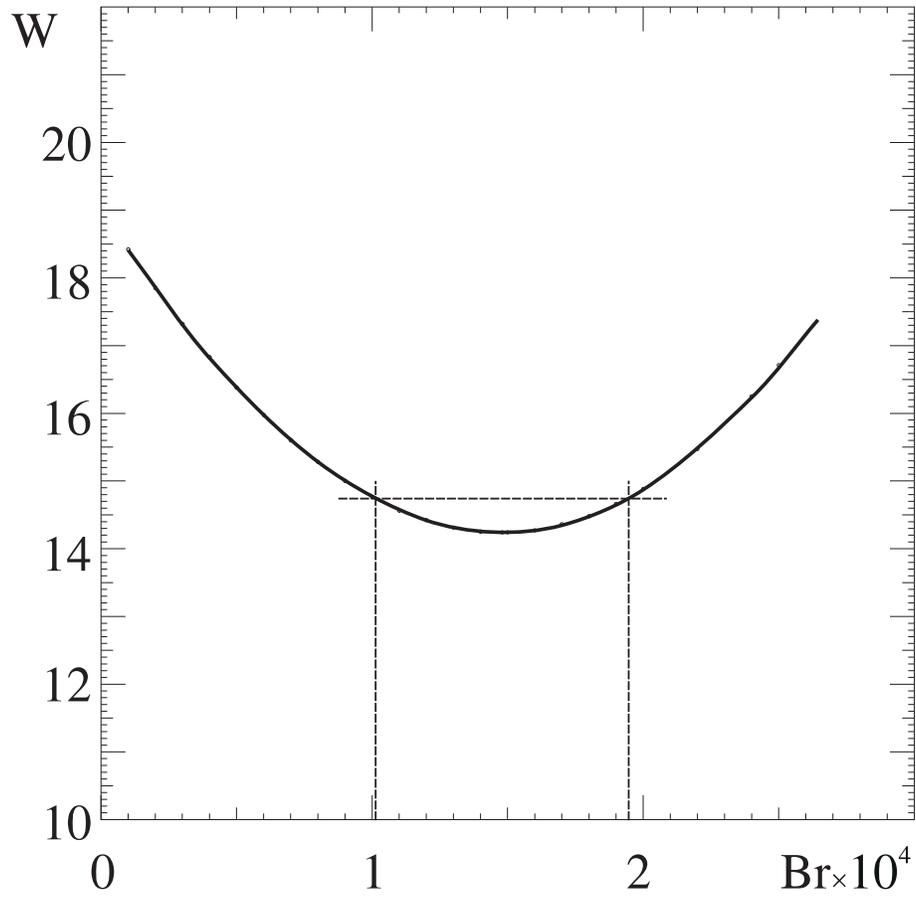,%
                  width=\textwidth}}
 \end{center}
 \caption{Dependence of maximum likelihood function on the 
         $BR(\phi\to\eta\pi ^0\gamma)$}
 \label{LENA6}
\end{figure}

The energy dependence of the visible
cross-section was fitted by the sum of process (\ref{tenth}), background process
$\phi \to \eta \gamma $ and non resonant contribution from the process $e^+e^- \to
\omega \pi^0$. The dependence of likelihood function on $B(\phi \to \eta\pi^0\gamma)$
is shown in fig.~\ref{LENA6}. The optimal value is 

\begin{displaymath}
B(\phi \to \eta\pi^0\gamma) = (1.3\pm 0.5)\cdot 10^{-4}
\end{displaymath}

\begin{figure}[htb]
 \begin{center}
 
 \vspace{-2cm}
 
  \hspace{-1cm}
  \mbox{\epsfig{figure=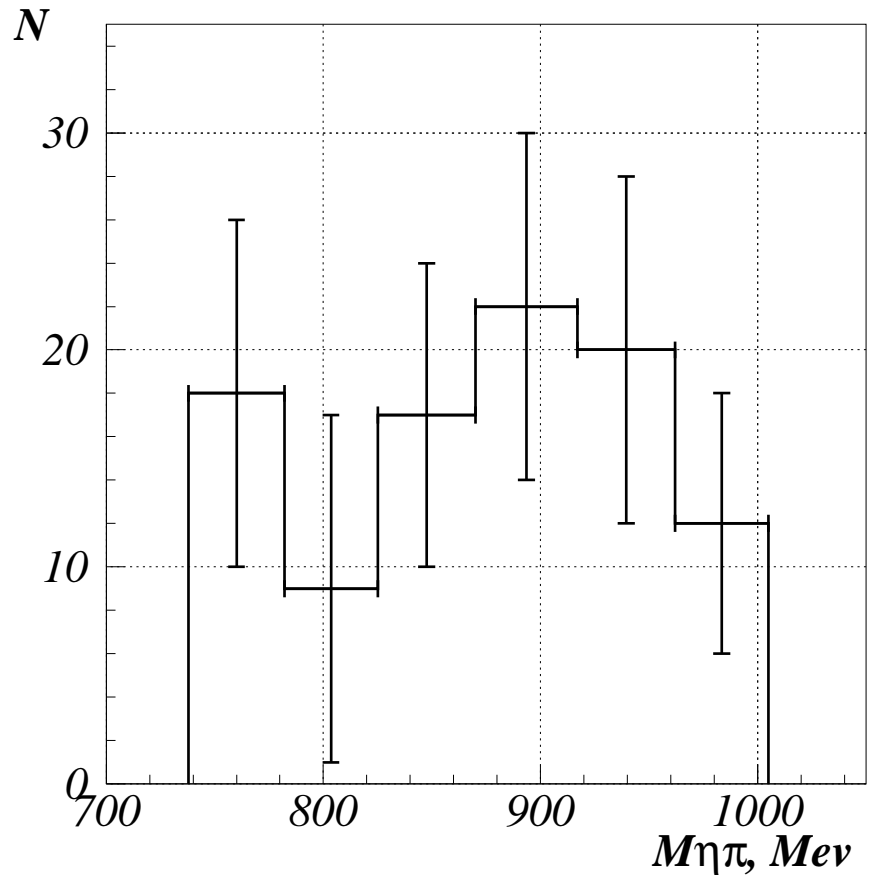,%
                   width=\textwidth}}
 \end{center}
 \caption{Measured mass spectrum of $\eta\pi ^0$ system.}
 \label{LENA7}
\end{figure}

The mass spectrum of $\eta\pi^o$-system is shown in
fig.~\ref{LENA7}

We also searched for $\phi \to \eta\pi^o\gamma$ decay in 
$\eta \to \pi^+\pi^-\pi^o$ and $\eta \to 3\pi^o$ decay modes,
but due to higher background and lower efficiencies only upper limits
for  $\phi \to \eta\pi^o\gamma$ branching ratio were obtained:

$B(\phi \to \eta\pi^0\gamma) < 1.5\cdot 10^{-4}$ $(\eta\to\pi^+\pi^-\pi^0)$,

$B(\phi \to \eta\pi^0\gamma) < 3\cdot 10^{-4}$ $(\eta\to 3\pi^0)$ \\
at $90\%$ confidence level.


\subsubsection{ Search for the $\phi \to K_S K_S \gamma $ process}

The $\phi \to K_SK_S\gamma$ decay also represents an electric
dipole transition, the width of which can be expressed as $M^2\omega^3$,
where $M$ is a matrix element and $\omega$ is a photon energy. The
$K_S K_S$ final state may be produced for example
in the decays of light scalar mesons $a_o(980)$ or $f_o(980)$.
The expected branching ratio of the $\phi \to K_SK_S\gamma$ decay,
obtained on the basis of the theoretical work \cite{U3} and table
values on $a_o(980)$ ט $f_o(980)$- mesons \cite{PDG} , can lie within
a wide range of $10^{-7} \div 10^{-9}$. Currently there are no available 
experimental data on this decay.

     The  search for $\phi \to K_SK_S\gamma$ decay was performed in the
decay mode $\phi \to K_SK_S\gamma \to 2\pi^o 2\pi^o \gamma \to 9\gamma$.
The events containing
8 or 9 photons were analyzed with the requirement, that invariant masses of
four different photon pairs in an event must be close to a mass of $\pi^o$
($m_{\gamma\gamma} = 110 \div 160~MeV$). Then was required, that of six
possible $\pi^o \pi^o$- pairs at least one has an invariant mass close to
that of $a_o$- meson ($430\div 560~MeV$). To suppress background from
$\phi \to \eta \gamma \to 3\pi^o \gamma \to 7\gamma$ decay, the energy
of the most energetic photon in an event was required to be less than
360~MeV. In addition the following constraints were applied:
$ E_{tot}/2E = 0.8 \div 1.05$;  $P_{tot}/2E < 0.2$;  $\chi ^2_E < 50$.

With such selection criteria the detection efficiency, obtained using
simulated events of the process under study, varied from 
$\varepsilon _1=0.4\%$ up to $0.7\%$, depending on the recoil photon energy
$\omega = m_\phi - m_{K_SK_S} = 0 \div 24~MeV$.
The SND threshold for photons is about 10~MeV.
The detection efficiency  with such a threshold is equal to $(0.7 \pm 0.1)\%$.
During processing of experimental data from the scans 2, 3, and 4,
total of  $N_o=16$ events were found. All of them could be attributed to a 
background  from $\phi \to K_SK_L$ decay. Thus, the result
can be expressed only in terms of an upper limit:
$B(\phi \to K_SK_S\gamma)
< 2\sqrt{N_o}/N_{\phi}\varepsilon _1
< 3.2\cdot 10^{-4}$, at 95\% confidence level.


\subsection{ Rare nonradiative decays of $\phi$- meson}
\subsubsection{ The process
$e^+ e^- \to \omega \pi^o \to \pi^+ \pi^- \pi^o \pi^o$ }

Reaction
\begin{equation}
\label{D1}
e^+e^- \to \pi^+ \pi^- \pi^o \pi^o
\end{equation}
was studied in the  $\phi$- resonance region earlier \cite{ND}.
This process proceeds purely via
$\omega \pi^o$- intermediate state and its cross section is close to  $10~nb$.
The main contribution comes from
$\rho (770)$- meson. In the $\phi$- peak region the possibility exists
to detect contribution from the $\phi \to \omega \pi^o$ decay, which
may reveal itself as a narrow interference pattern on a smooth energy
dependence of the $\rho (770)$ contribution. Probability of
this decay was estimated to be at a level of  $5\cdot 10^{-5}$
\cite{DRU2}. 

For analysis the events were selected with two charged particles
and four photons. The energy-momentum conservation 
and presence of two $\pi^o$- mesons ($\chi^2_M < 30$) were required.
To suppress background from the process $e^+e^- \to K^+K^-$,
restrictions were imposed on specific ionization losses in the drift
chamber.
Background from the $\phi \to \eta \gamma \to \pi^+ \pi^- \pi^o \gamma$
decay was rejected by a requirement, that the most energetic photon in
the event must have an energy less  than 360 MeV. The 
$\phi \to \pi^+ \pi^- \pi^o $ decay can also fake the process
under study if some stray photons are detected. To suppress it, an
additional requirement was imposed: if the invariant mass of the two most
energetic photons is close to that of $\pi^o$- meson, then invariant mass of
the other pair must differ from it.

\begin{figure}[htb]
  \begin{center}
    \mbox{\epsfig{figure=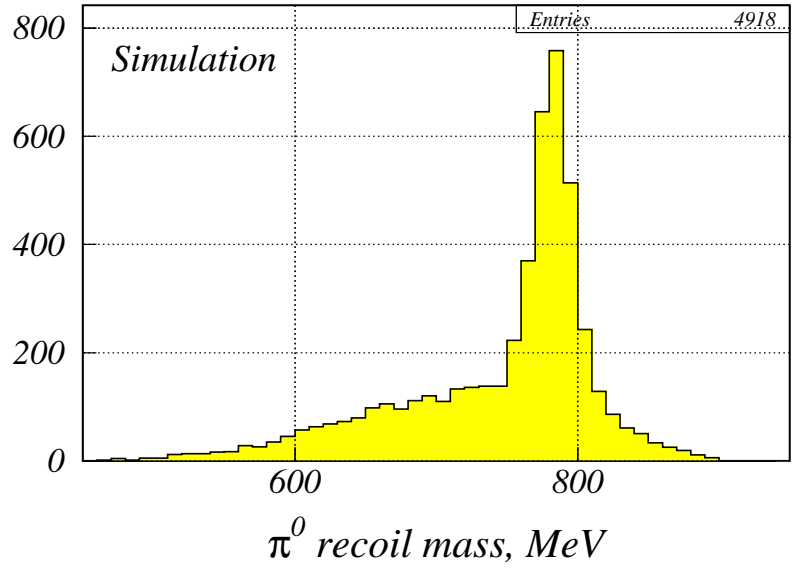,%
                  width=\textwidth}}
        \vskip 2cm
    \mbox{\epsfig{figure=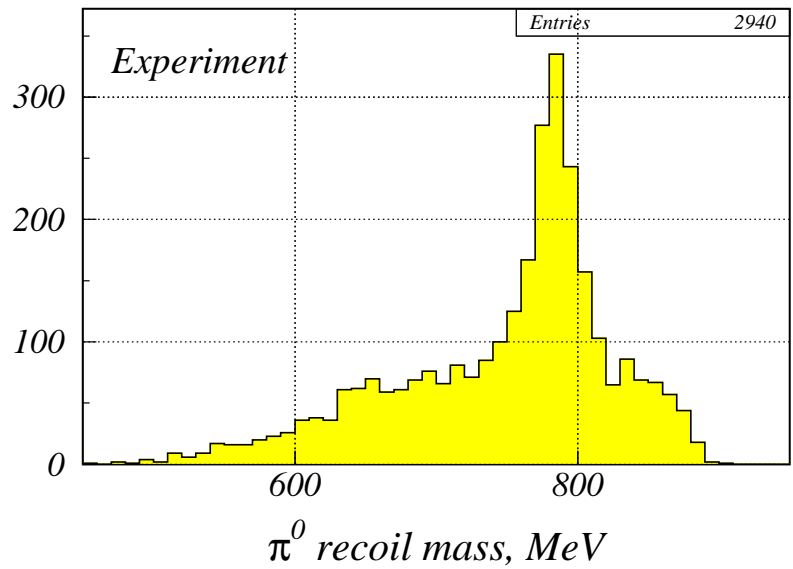,%
                  width=\textwidth}}
  \end{center} 
\caption{Recoil mass spectrum of $\pi^o$- mesons in the
$e^+e^- \to \pi^+ \pi^- \pi^o \pi^o$ process. 
Top --- simulation; 
bottom --- experiment.}
\label{DRU_MRECPI}
\end{figure}

Event detection efficiency under these selection criteria,
estimated using simulation, is equal to $\varepsilon = 20\%$.
In  the recoil mass spectrum of $\pi^o$- mesons,
shown in Fig.~\ref{DRU_MRECPI},
the $\omega$- meson peak is clearly seen. On the other hand,
presence of a nonresonant contribution is an evidence of some
background with intermediate states other than $\omega \pi^o$.
To account for this background the special subtraction procedure was
developed. It is based on background estimation in the kinematic
regions not containing the process under study.

The energy dependence of the cross section was approximated according
to the formula (\ref{CS_INTERF}),
with $Z$, the complex interference amplitude, varying, according to
theory, in the limits:
$Z=[(11\div 34) - i (11\div 24)]\cdot 10^{-2}$,
and decay probability, described by an expression:

$B(\phi \to \omega \pi^o) = \sigma _o(m_{\phi}) |Z|^2/
\sigma _{\phi}$.

\begin{figure}[htb]
  \begin{center}
    \mbox{\epsfig{figure=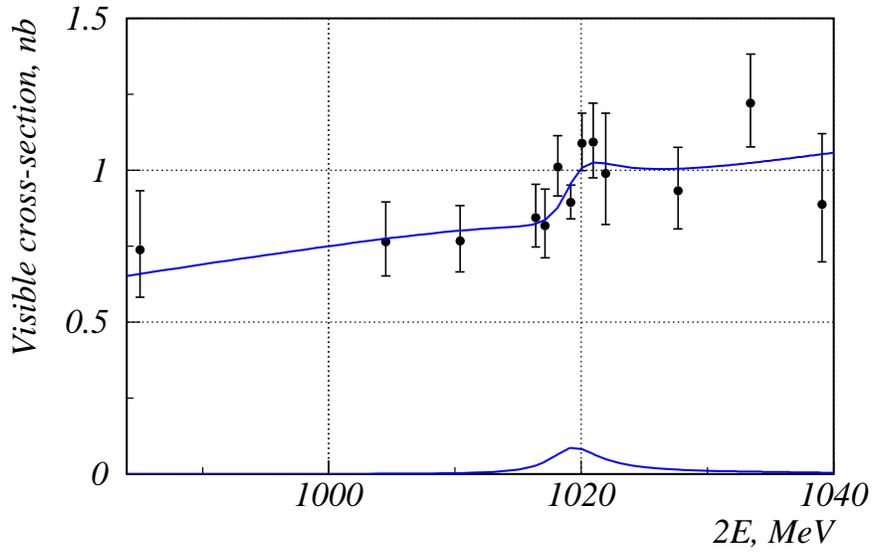,%
                  width=\textwidth}}
  \end{center} 
\caption{Detection cross section and fit parameters for the
$e^+e^- \to \omega \pi^o \to \pi^+ \pi^- \pi^o \pi^o$ process.}
\label{DRU_CS}
\end{figure}

For approximation the detection cross section was expressed as in
(\ref{CROSSSECTION}), and the result of the procedure is shown in
Fig.~\ref{DRU_CS}. The following values of parameters were obtained:

$\sigma _o = 6.4 \pm 0.5 \pm 1.0~nb$ -- reaction (\ref{D1})  cross section 
via
$\omega \pi^o$- intermediate state,

$Re Z =(10\pm 7)\cdot 10^{-2}$;
$Im Z =(0\pm 6)\cdot 10^{-2}$;
$\chi^2/N_D = 7/9$.

Systematic error of  $\sim 15\%$ is caused by inexact simulation
and background subtraction error. As could be seen from Fig.~\ref{DRU_CS}
and the values presented above, it is possible to establish only
the following upper limit
of the decay branching ratio:

 $B(\phi \to \omega \pi^o) < 5\cdot 10^{-5}$
at $90\%$ confidence level.

The contribution of non- $\omega \pi^o$ intermediate states is a subject
of further investigation.

\subsubsection{ The $e^+e^- \to \mu ^+ \mu ^- $ process
near $\phi$- resonance}

Measurements of lepton widths of light vector mesons $\rho$, $\omega$,
$\phi$ at electron-positron colliders were usually performed through
summation of cross sections over all decay channels.
Study of the process 
\begin{equation}
e^+e^- \to \mu ^+ \mu ^-
\label{BUR1}
\end{equation}
in the vicinity of the $\phi$- peak gives the possibility of direct measurement
of 
$B_{\mu \mu} \equiv B(\phi \to \mu ^+ \mu ^-) $.
Expected value of
$B_{\mu \mu}$, calculated from $\mu - e$- universality, is equal to:
\begin{equation}
B_{\mu \mu} \simeq B_{ee} = (3.00 \pm 0.06)\cdot 10^{-4}
\label{BUR2}
\end{equation}

The world average value, taken from PDG (1996) \cite{PDG}:\\
$B_{\mu \mu}= (2.48\pm 0.34)\cdot 10^{-4}$.

The cross section of the process (\ref{BUR1}) in the vicinity of the 
$\phi$- resonance can be expressed (in linear approximation) in
the form  (\ref{CS_INTERF}), where
$Z=3\sqrt{B_{\mu \mu}B_{ee}} e^{i\beta } / \alpha $, and
$\beta$ is an interference phase.

Events were selected according to following conditions:
$N_{cp}=2$; $N_{\gamma}=0$; $\Delta \vartheta <20^o$;
$\Delta \varphi <10^o$; $|Z_{1,2}|<8~cm$; $R<1~cm$;
$\vartheta _{min} >45^o$.

The main sources of background were the following:

1) $e^+e^- \to e^+e^-$, was rejected using $e/\pi$- separation;

2) $e^+e^- \to \pi^+\pi^-$, was suppressed using the outer
anticoincidence system of the detector.

3) cosmic muons. These events are not peaked in time, measured by
outer scintillation counters, with respect to a
beam crossing. On the contrary, the events of the process (1)
are strongly peaked with a RMS of $\sigma \sim 1 ns$. This permits to
reject main part of cosmic muons. Remaining background is suppressed
by limiting maximum distance from charged particle tracks
to the beam axis.

The detection efficiency for the process (1) is described by the
following formula:

$\varepsilon_{\mu \mu} = \varepsilon_{sel} \cdot
\varepsilon_{\vartheta} \cdot \varepsilon_{up} \cdot
\varepsilon_{\mu} \cdot \varepsilon_{FLT} \simeq 0.32$,

where ${\varepsilon}_{sel} = 0.75$ is a detection efficiency
in the conditions, listed above;

$\varepsilon_{\vartheta} = 0.55$ is an acceptance in polar plane;

$\varepsilon_{up} = 0.99 $ is the efficiency of anticoincidence
counters;

${\varepsilon}_{\mu} = 0.82$ probability for a muon to trigger
outer system;

$\varepsilon_{FLT} = 0.97 $ in the second scan, and
0.91 in the third and fourth scans --- first level trigger efficiency.

\begin{figure}[htb]
  \begin{center}
    \mbox{\epsfig{figure=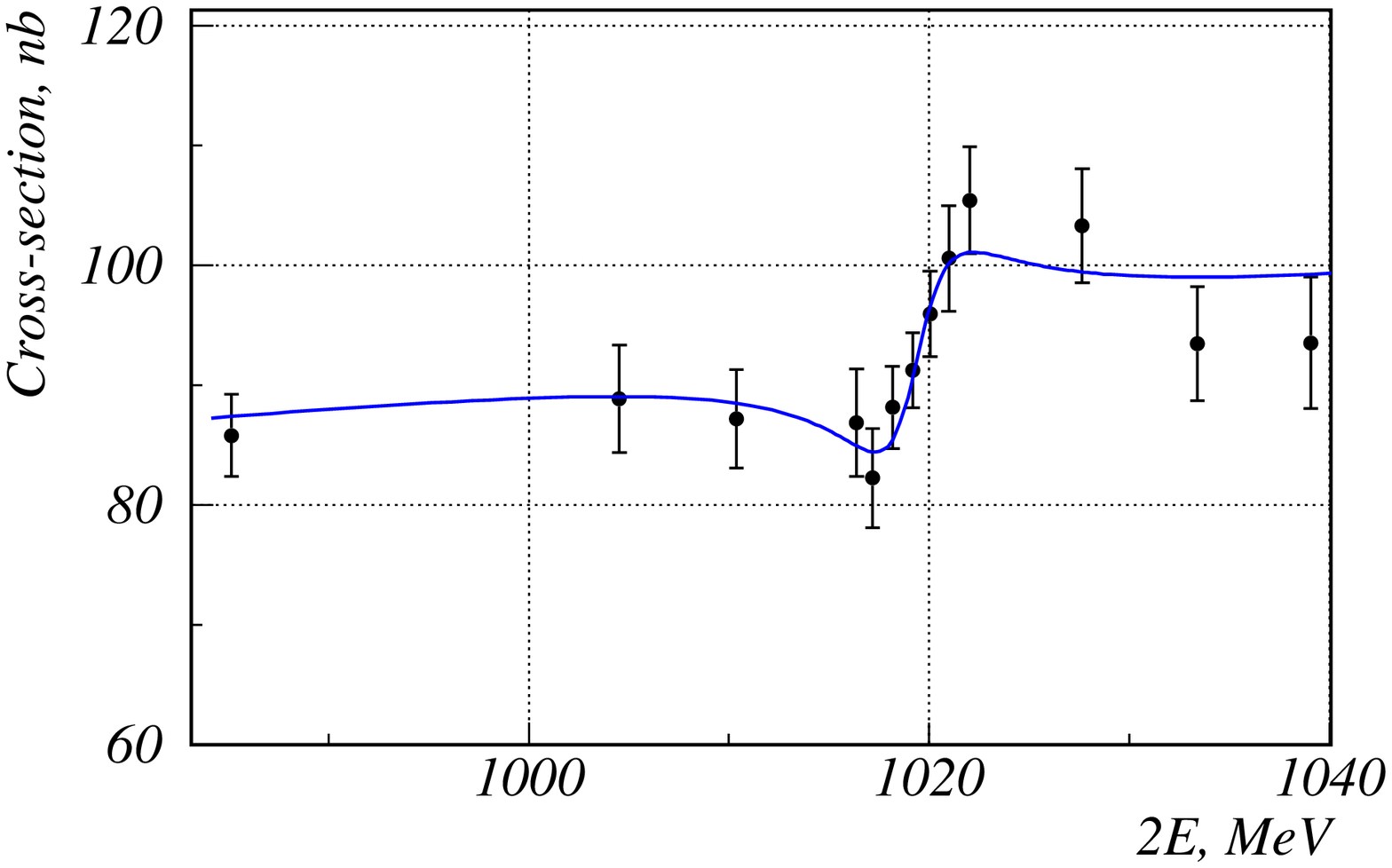,%
                  width=\textwidth}}
  \end{center}
\caption{$e^+e^- \to \mu^+ \mu^-$ cross section.}
\label{BUR_CS}
\end{figure}

At first, fit was performed with an interference phase as a free
parameter. The result was $\beta = 0.13 \pm 0.11$. Then the value of $\beta$ 
was set to zero and the fit was repeated. The results are shown
in Fig.~\ref{BUR_CS} and in Table~\ref{TAB_BUR}.

\begin{table}[!htb]
\caption{The results of cross section fitting for the 
$e^+e^- \to \mu ^+ \mu ^- process$
}
\label{TAB_BUR}

\begin{center}

\footnotesize

\begin{tabular}{||c|c|c|c|c||}

\hline

Experiment      & $\sigma _o$,nb & $\sigma ',10^{-2} MeV^{-1}$ &
$B_{\mu \mu}, 10^{-4}$ & $\chi^2/N_{D}$\\
name  &                &   &                         &   \\
\hline
FI9602        & 96.7$\pm$1.6&-0.05$\pm$0.21 &2.8$\pm$1.4 & 9.2/6 \\
FI9604        & 95.9$\pm$1.2& 0.13$\pm$0.12 &2.2$\pm$0.9 & 12.2/9\\
FI9602\&FI9603\&FI9604
              & 96.7$\pm$0.9& 0.09$\pm$0.09 &2.4$\pm$0.8 & 29/26 \\
\hline
\end{tabular}
\end{center}
\end{table}
The errors indicated in the Table~\ref{TAB_BUR} are all statistical.
Possible sources of additional systematic errors are contributions
from other decays of $\phi$- meson (for example:
$\phi \to K^+K^-, \rho \pi,...$); systematic errors in simulation,
inadequate treatment of radiative corrections. In this preliminary
analysis the systematic error in $\sigma _o$ was estimated to be 
about $5\%$ and the error in $B_{\mu \mu}$ --- 20\%.
Let us finally list the main results: 

$\sigma _o = 96.8 \pm 0.9 \pm 5.0~nb$,

$B_{\mu \mu} = (2.4 \pm 0.8 \pm 0.6)\cdot 10^{-4} $,

$B_{e \mu} = \sqrt{B_{ee}\cdot B_{\mu \mu}} =
               (2.74 \pm 0.44 \pm 0.6)\cdot 10^{-4}$.

\subsection{ Rare decays of $\eta(550)$- meson}
\subsubsection{ Upper limit of the $\eta \to 2 \pi^o$ decay }

CP-violation is one of the most intriguing puzzles in the particle physics.
Up to now the only place, when CP-violation was
observed was a neutral kaons system. Standard Model (SM) predicts much more
pronounced effects in decays of $B$- mesons.

Flavor conserving CP- violating effects, like $\eta \to 2\pi $ decay,
within a framework of SM are very weak:
$B(\eta \to \pi^+ \pi^-) < 2 \cdot 10^{-27}$ \cite{ZUR1,ZUR2}.
But some other, different from SM, models exist in which such effects are
many orders of magnitude larger.
$B(\eta \to \pi^+ \pi^-) < 10^{-15} \div 10 ^{-16} $
\cite{ZUR2,ZUR3,ZUR4}.
In any case, search for such unusual effects can produce interesting and
unexpected results.

The only existing experimental upper limit ---
$B(\eta \to \pi^+ \pi^-) < 1.5 \cdot 10^{-3}$ was reported in the work
\cite{ZUR8}. Recently this result was confirmed by CMD-2 detector:
$B(\eta \to \pi^+ \pi^-) < 2 \cdot 10^{-3} \cite{CMD2_PROCEED}$.
In this work search for the process
$\eta \to \pi^o \pi^o$, with $\eta$-s produced in the reaction
$e^+e^- \to \phi \to \eta \gamma $ was carried out.

\begin{figure}[htbp]

\vspace{2cm}

\epsfxsize=0.49\textwidth
  \begin{minipage}[t]{0.49\textwidth}
  \centerline{\epsfbox{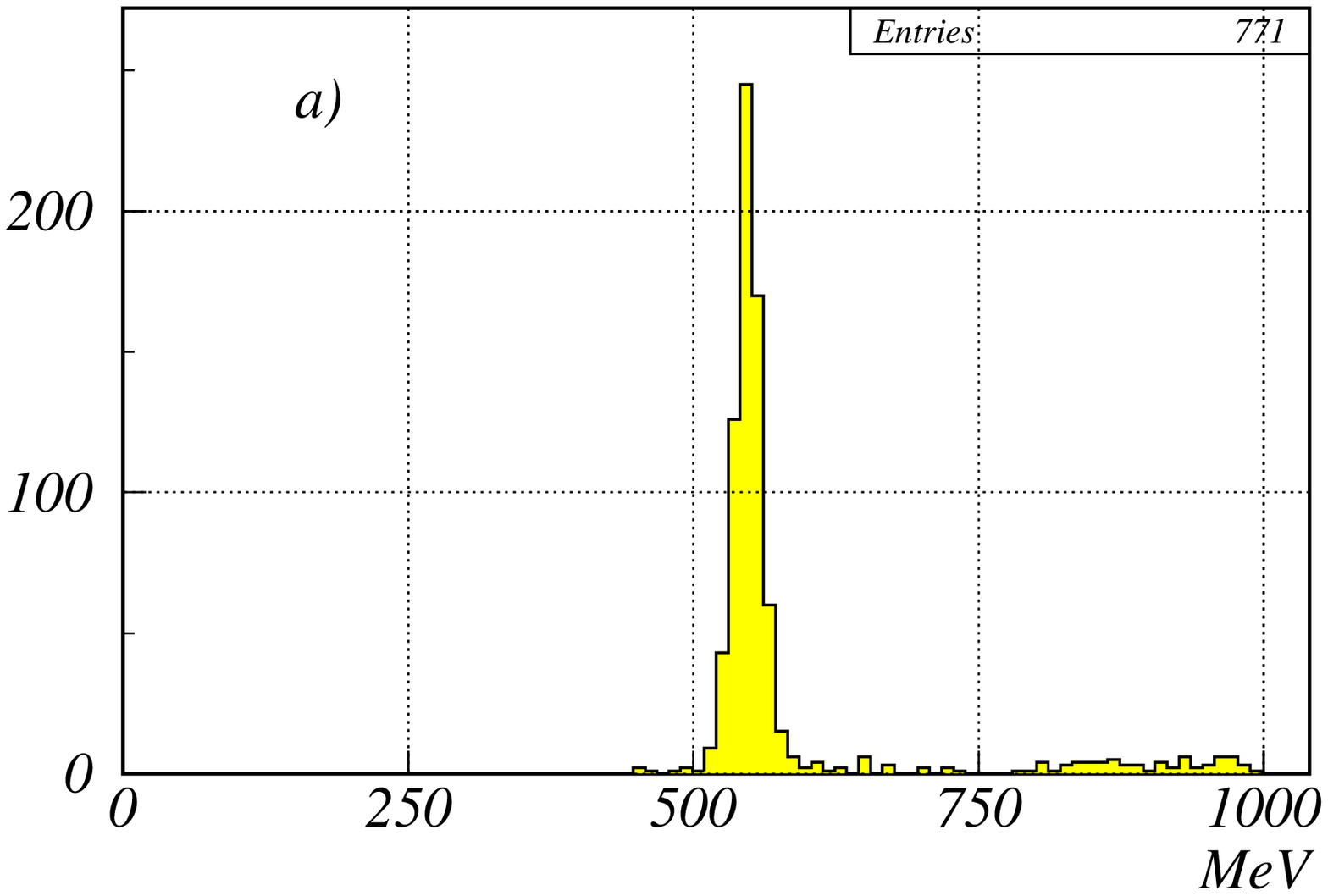}}
  \end{minipage}
\hfill
\epsfxsize=0.49\textwidth
  \begin{minipage}[t]{0.49\textwidth}
  \centerline{\epsfbox{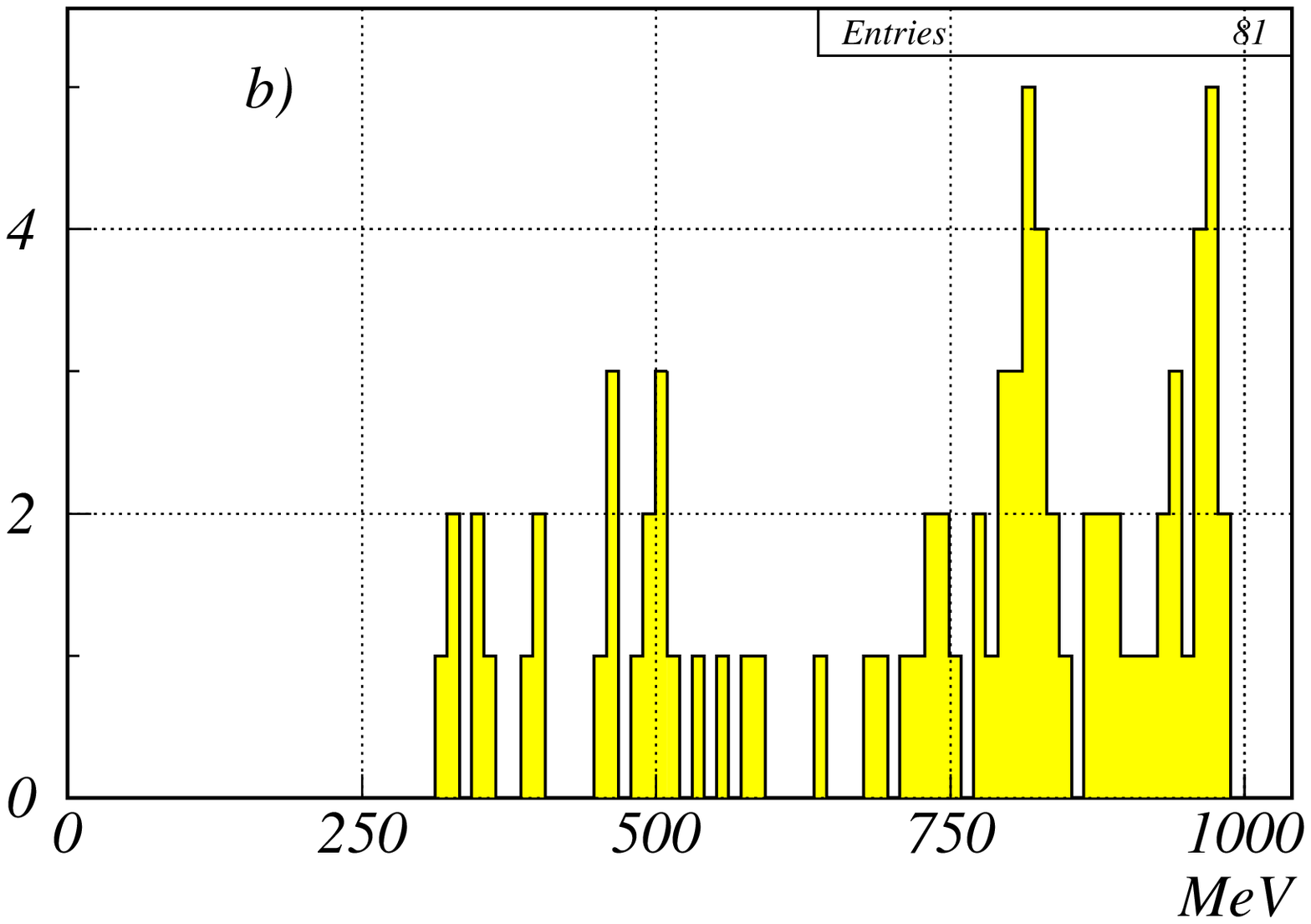}}
  \end{minipage}
\hfill
\epsfxsize=0.49\textwidth
  \begin{minipage}[t]{0.49\textwidth}
  \centerline{\epsfbox{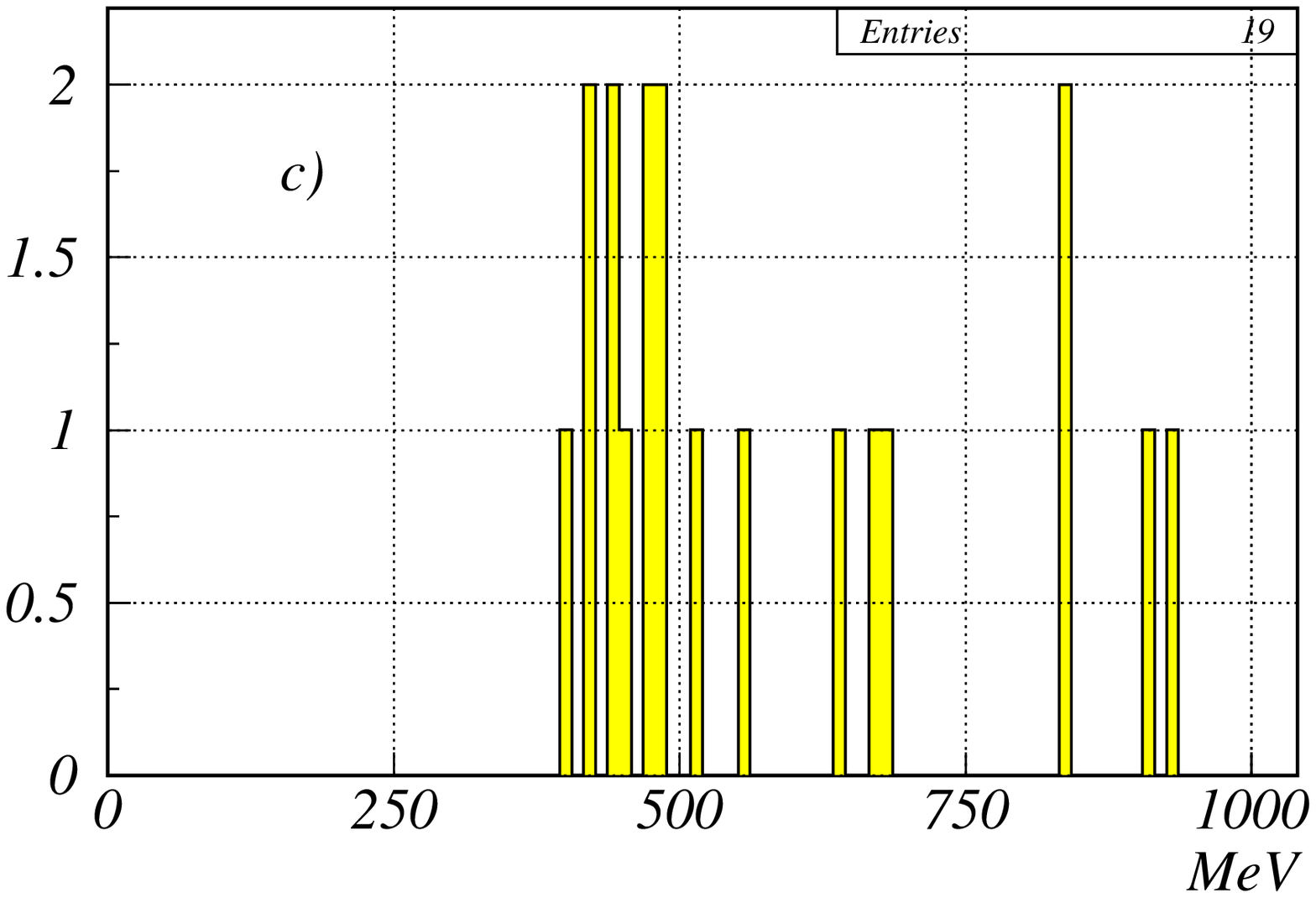}}
  \end{minipage}
\hfill
\epsfxsize=0.49\textwidth
  \begin{minipage}[t]{0.49\textwidth}
  \centerline{\epsfbox{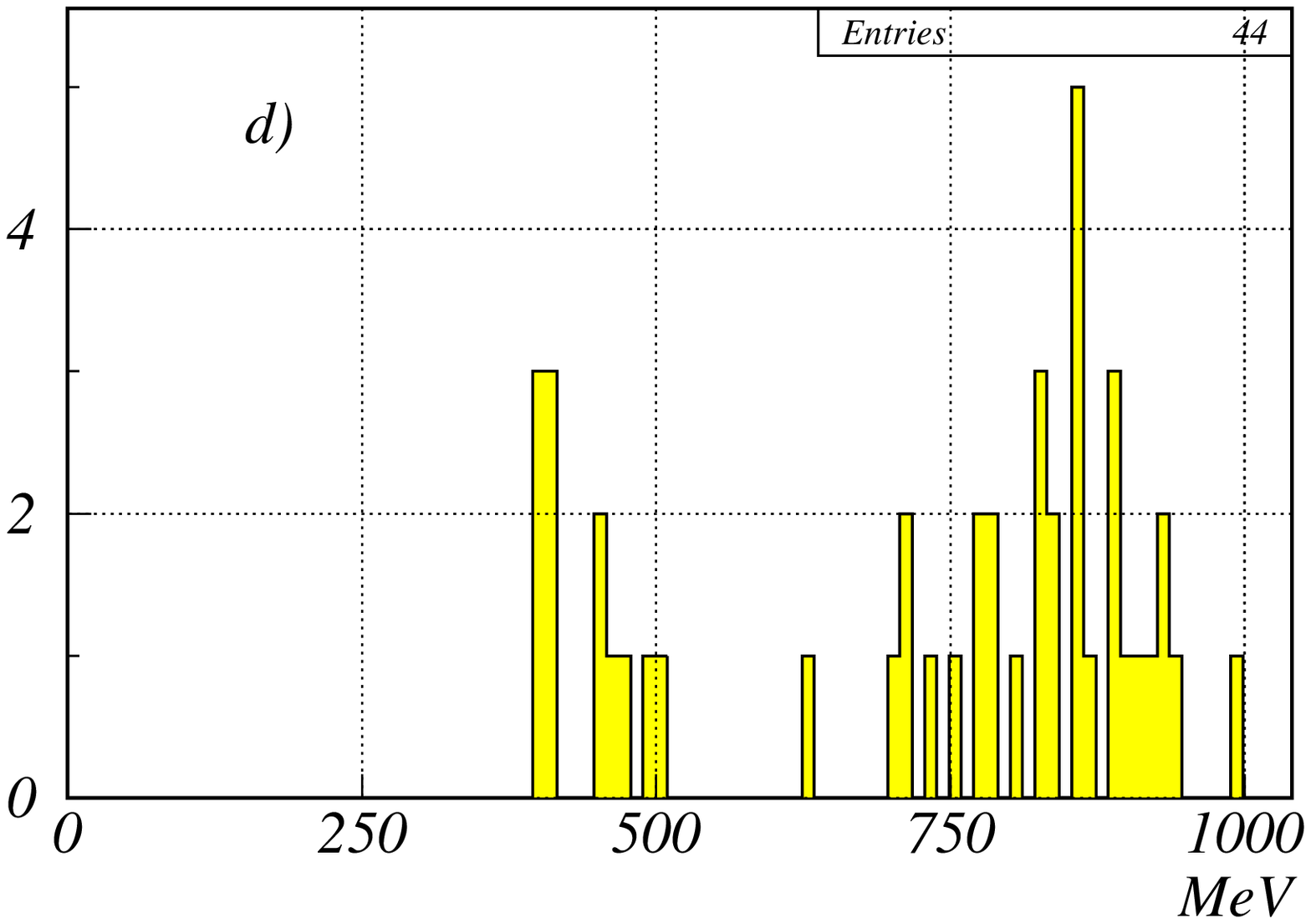}}
  \end{minipage}
\caption{ \label{FIG_ZUR}
Photon recoil mass spectrum in the $e^+e^- \to \varphi
\to \pi^o \pi^o \gamma$ process;
a --- $\eta \to 2\pi^o$ simulation;
b --- $\omega \pi^o$  $ (\omega \to \pi^o \gamma)$ simulation;
c --- $\eta \to 3\pi^o$ simulation;
d --- SND experiment.}
\end{figure}

Events for analysis were selected with two  $\pi ^o$- mesons and a photon,
satisfying energy--momentum conservation. The estimated detection efficiency
is equal to 8.6\%. In Fig.~\ref{FIG_ZUR} photon recoil mass spectra in the
process $e^+e^- \to \phi \to \eta \gamma \to \pi^o \pi^o \gamma $
are shown.  Fig.~\ref{FIG_ZUR}a depicts an expected spectrum in the
process being searched for; Figs.~\ref{FIG_ZUR}b,c show spectra for the
main background processes, and Fig.~\ref{FIG_ZUR}d --- experimental data.
Unfortunately the number of simulated events is not yet sufficient to determine
contribution from $\phi \to K_S K_L $ (statistics at least equal to
the experimental one is needed, i.e. $\sim 4 \cdot 10^{6}$ $\phi$- mesons).
As it is seen in Fig~\ref{FIG_ZUR}d, there is no indication of the
$\eta \to \pi^o \pi^o$ decay and thus, only upper limit can be placed:

$B(\eta \to \pi^o \pi^o) < 6 \cdot 10^{-4}$ at 90\% confidence level.

\subsection{ Nonresonant processes in the $\phi$- resonance region}

\subsubsection{ The $e^+ e^- \to 3 \gamma$ reaction}

This process is interesting from the point of view of QED testing
near $\phi$- resonance and as an important source of background in the
studies of the decays $\phi \to \eta \gamma, \pi^o \gamma \to 3\gamma $.

Events were selected according to following criteria:
          $N_\gamma=3$; $\vartheta_{min}>27^o$;
$0.75<E_{tot}/2E<1.1$; $P_{tot}/2E<0.1$;
$\chi ^2_{\gamma}<15$;  $\chi ^2_E < 30$.

\begin{figure}[htb]
  \begin{center}
    \mbox{\epsfig{figure=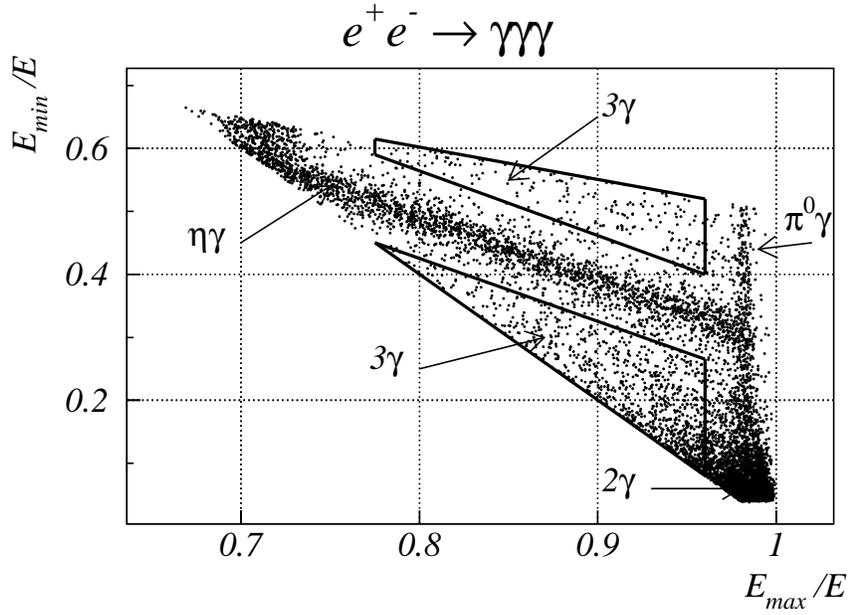,%
                  width=\textwidth}}
  \end{center}
\caption{Dalitz plot for the events of the
$e^+e^- \to 3\gamma$. process}
\label{ARTUR_DALITC}
\end{figure}

\begin{figure}[htb]
  \begin{center}
    \mbox{\epsfig{figure=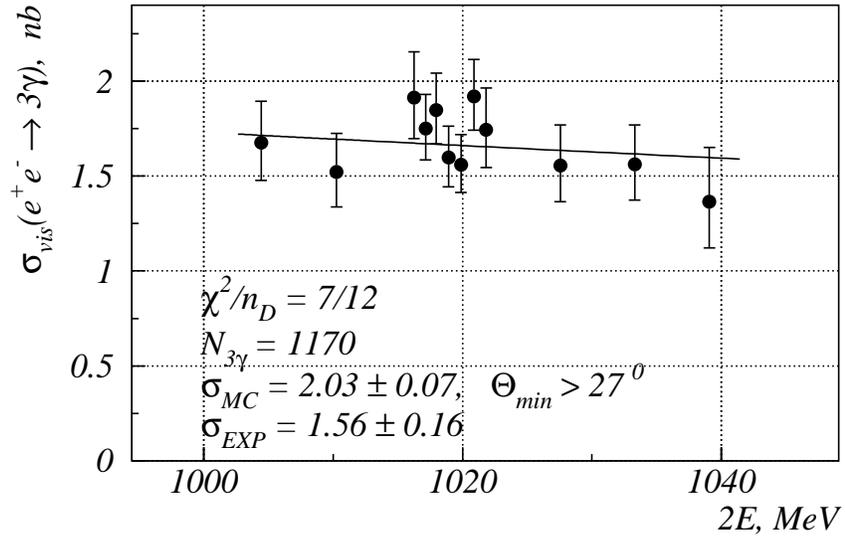,%
                  width=\textwidth}}
  \end{center}
\caption{ $e^+e^- \to 3\gamma$ detection cross section.}
\label{ARTUR_CS}
\end{figure}

\begin{figure}[htb]
  \begin{center}
    \mbox{\epsfig{figure=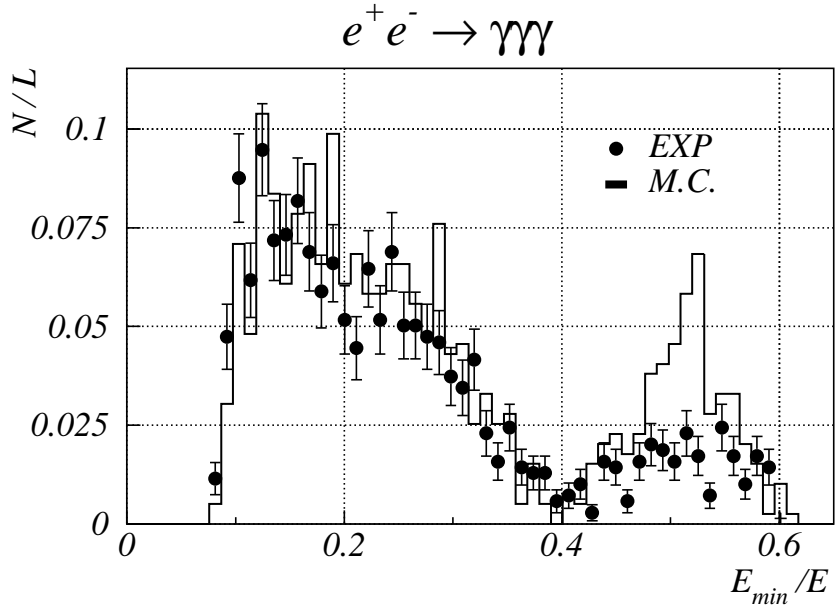,%
                  width=\textwidth}}
  \end{center}
\caption{Lowest energy photon spectrum in 
$e^+e^- \to 3\gamma$ events.}
\label{ARTUR_SPECTR}
\end{figure}

The Dalitz plot for  the selected events is shown in Fig.~\ref{ARTUR_DALITC}.
Regions of $\phi$- meson decays $\phi \to \eta \gamma ,
\pi^o \gamma  \to 3 \gamma $, are marked and events from them were
excluded from further analysis.
The energy dependence of the visible cross section for the selected events
is shown in Fig.~\ref{ARTUR_CS}. The estimated cross section at 510~MeV 
is equal to $\sigma _{MC} = 2.03 \pm 0.07~nb$, while the experimental
value is 20\% smaller:  $\sigma _{exprm} = 1.56 \pm 0.16 \pm 0.20~nb$.
The cause of such discrepancy is under study.
In Fig.~\ref{ARTUR_SPECTR} the spectrum of the least energetic photon
in an event is presented. One can see, that the largest difference between
theory and experiment lies in the region $E_{min}/E \sim 0.5$.

 The $e^+ e^- \to 3 \gamma$ process is a source of background for
neutral decays of vector mesons and, on the other hand, its cross section is
precisely known. Its investigation is important for understanding of systematic
errors in other neutral processes and accuracy of luminosity estimations.


\subsubsection{ The process $e^+e^- \to e^+e^-\gamma$ and
search for $\eta \to e^+ e^- $ decay }

The process $e^+e^- \to e^+e^-\gamma$ is important for us due to
several reasons:

\begin{enumerate}
\item this process can be used for testing of QED and
should be taken into account as a correction for 
Bhabha scattering, in order to achieve accuracy of 
luminosity measurements of $\leq 1\%$.

\item this process is a background to different hadronic processes,
i.g. $\phi \to \pi^o \gamma, \eta \gamma \to e^+e^-\gamma$.
\end{enumerate}

This process was studied earlier in several experiments: with
OLYA detector \cite{OLQ} with an integrated luminosity of 
$\Delta L =1.5~pb^{-1}$, at ADONE collider  \cite{ADONE}
with $\Delta L =0.12~pb^{-1}$.

The results of this work are based on data corresponding to integrated
luminosity of $1.8~pb^{-1}$. During data processing the subset of
events with Dalitz pairs was also included. In such events the invariant mass
of  $e^+e^-$- pair is very small and the whole process looks like a
two-photon annihilation in which one of photons is slightly out of a
mass shell, and it subsequently decays into $e^+e^-$- pair.

The $e^+e^-\gamma$ final state is favorable for the search of the decay
$e^+e^- \to \phi \to \eta \gamma \to e^+e^-\gamma$.
The $\eta \to e^+e^-$ decay is strongly suppressed in comparison with
$\eta \to \gamma \gamma$ by a helicity factor $(m_e/m_{\eta})^2$ and
additional QED suppression factor $\alpha ^2$. As a result the expected
branching ratio is about $5\cdot 10^{-9}$ \cite{TANYA4}. The similar
decay $\pi ^o \to e^+e^-$ was observed at a level of  $\sim 10^{-7}$.
Current upper limit of the $\eta \to e^+e^-$ is equal to
$2\cdot 10^{-4}$ \cite{TANYA5}

Event selection was done using the following main criteria:
$N_{cp}=2$; $N_\gamma=1,2$; $\vartheta_{min}>36^o$;
$R<0.5 cm$; $|Z|<10 cm$; $E_{tot}/2E>0.8$; $P_{tot}/2E<0.15$;
$\Delta \varphi _{ee}> 5^o$; $\chi ^2_E < 40$;
$\Delta \Omega _{ee}< 50^o$ --- spatial angle between
particles for Dalitz pairs.

\begin{figure}[htb]
  \begin{center}
    \mbox{\epsfig{figure=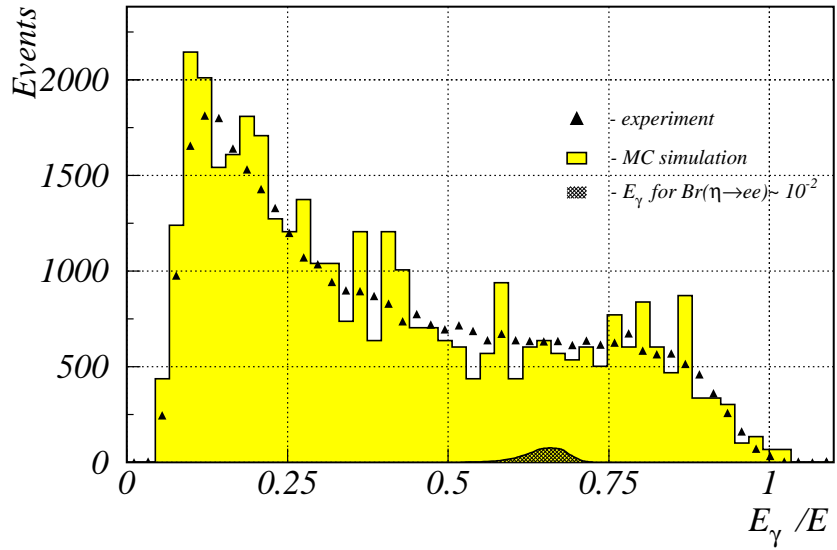,%
                  width=\textwidth}}
  \end{center}
\caption{Photon spectrum in the $e^+e^- \to e^+e^-\gamma$ process.
Peak at $E_{\gamma}=0.7 E$ comes from the possible
$\eta \to e^+e^-\gamma$ decay with
a branching ratio close to 1\%.}
\label{TANYA_SPECTR}
\end{figure}

\begin{figure}[htb]
  \begin{center}
    \mbox{\epsfig{figure=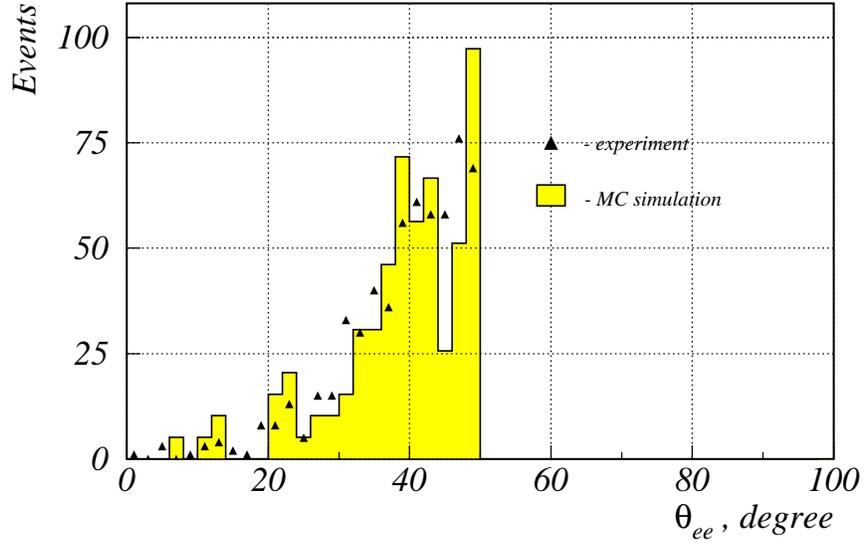,%
                  width=\textwidth}}
  \end{center}
\caption{ Distribution over the spatial angle between electrons in the
$e^+e^- \to e^+e^-\gamma$ process.}
\label{TANYA_ANGLE}
\end{figure}

Fig.~\ref{TANYA_SPECTR} shows the photon energy spectrum and
in Fig.~\ref{TANYA_ANGLE} the distribution over angle between
electrons is shown. There is also a  region indicated, where the
angle between electrons is very small (Dalitz pairs).
The number of events in  this region is in agreement with an expected one.

Energy dependence of the $e^+e^- \to e^+e^-\gamma$ cross section was
approximated using the following expression:

$\sigma _{vis}(E) = \varepsilon _1\sigma _o(\frac{E_o}{E})^2
+ \varepsilon \sigma _{3\pi}$.

The first term is responsible for the process under study, while the
second describes background from decays of  $\phi$- meson, especially 
$\phi \to 3\pi$.
It was estimated from background simulation, that 
$\varepsilon = (0.10 \pm 0.03)\%$.
The $\varepsilon _1$ coefficient was determined as a ratio between
the number of simulated events, satisfying selection criteria and the total
number of simulated events. The simulation was conducted with the following
constraints: $\vartheta _{min} > 27^o$, $E_{\gamma} > 5~MeV$, and
corresponding total cross section under such constraints is equal to
$\sigma _o =570 \pm 15~nb$, and $\varepsilon _1$ was found to be
equal to  $(3.5 \pm 0.1)\%$;

The following results were obtained for approximation parameters:
$\sigma _o =550 \pm 10~nb$; $\varepsilon = (0.08 \pm 0.05)\%$;
$\chi^2/N_D = 10/11$.
Only statistical errors are indicated. Additional systematic error is
estimated as 10\%.

The  $\eta \to e^+e^-$ decay could reveal itself as a peak in the photon spectrum
in Fig.~\ref{TANYA_SPECTR} at $E_\gamma = 360~MeV$. The detection efficiency,
obtained from simulation is $\varepsilon _2=(40\pm 1)\%$. After fitting of the
spectrum the following upper limit was obtained:
$B(\eta \to e^+e^-)<9\cdot 10^{-4}$. Although this limit is higher than
that of PDG Table, increase in statistics and taking into account the energy
dependence of the spectrum shape will permit us to improve this result.


\subsubsection{ $e^+ e^- \to \omega \pi^o \to
        \pi^o \pi^o \gamma$ }

There are several reasons of interest to the $e^+ e^- \to \omega \pi^o$
process. First, in this process may manifest themselves radial excitations of $\rho$- 
meson. Second, this process contributes into the
total cross section of $e^+ e^-$- annihilation into hadrons. Third,
its cross section is connected with the branching ratio of 
$\tau$- lepton decay into   $\omega \pi \nu _{\tau}$.

Within the Vector Dominance Model (VDM) cross section of this process is
determined by $g_{\rho \omega \pi }$ constant, which also controls
other processes: $\omega \to 3\pi, \pi ^o \gamma$; $\rho \to \pi ^o \gamma$;
$\pi ^o \to 2\gamma$, etc.

In our case the possibility arises to measure this constant at 
$\sqrt{s} =1~GeV$, which can help to determine boundaries of
applicability of simple VDM model.

The process $e^+ e^- \to \omega \pi^o$ in the energy region $2E>1~GeV$
was studied earlier in the experiments with ND detector at VEPP-2M in
two main decay modes of  $\omega $- meson:
$\omega \to \pi ^o \gamma$ \cite{ZURO1} and
$\omega \to \pi ^+ \pi ^- \pi ^o$ \cite{ND,ZURO2}.

Some indirect results on this reaction were extracted from  $\tau $- 
lepton decay spectra \cite{ND,ZURO3}.

In this paper data on the process 
$e^+ e^- \to \omega \pi^o \to \pi^o \pi^o \gamma $ in the energy range
$2E=1000 \div 1034~MeV$ are presented. Results are based on experimental
statistics from FI96 SND experiment corresponding to
integrated luminosity of $1.9~pb^{-1}$ or 4~million $\phi$- mesons 
produced.

The main event selection criteria are:
$N_{\gamma}=5$,
$E_{tot}/2E=0.75 \div 1.25$,
$\vartheta _{min} > 27^o$,
$P_{tot}/2E<0.1$.

Then additional requirements were imposed:
$\chi ^2 _M <40$;
$\chi ^2 _{\gamma} < 20$;
two photon pairs with $m_{\gamma\gamma} = m_{\pi ^o} \pm 30~MeV$
which could be produced by intermediate  $\pi ^o$- mesons must be 
found in the event;
$720 < m_{\pi ^o\gamma} < 840~MeV$.

Selected events originate mainly from the process under study:
$e^+ e^- \to \omega \pi^o \to \pi^o \pi^o \gamma $.
However they contain some background from the decay
$\phi \to \eta \gamma \to 3\pi ^o \gamma$.
The detection efficiency was estimated by simulation. For the
process under study it is equal to $\varepsilon _1= 21\%$ and
for the background  --- $\varepsilon _2= (1.8 \pm 0.1) \cdot 10^{-3}$.
As for another background source:
$\phi \to K_S K_L \to neutral$ $particles$, the available
simulation statistics is not sufficient to estimate its
contribution (minimum $10^6$ events are required, while only 
$6 \cdot 10^4$ are currently available).

\begin{figure}[htb]
  \begin{center}
    \mbox{\epsfig{figure=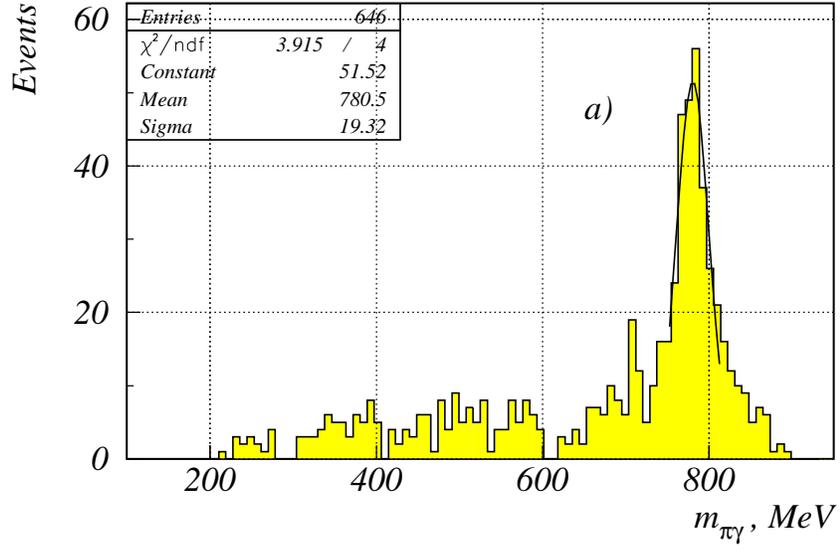,%
                  width=\textwidth}}
        \vskip 2cm
    \mbox{\epsfig{figure=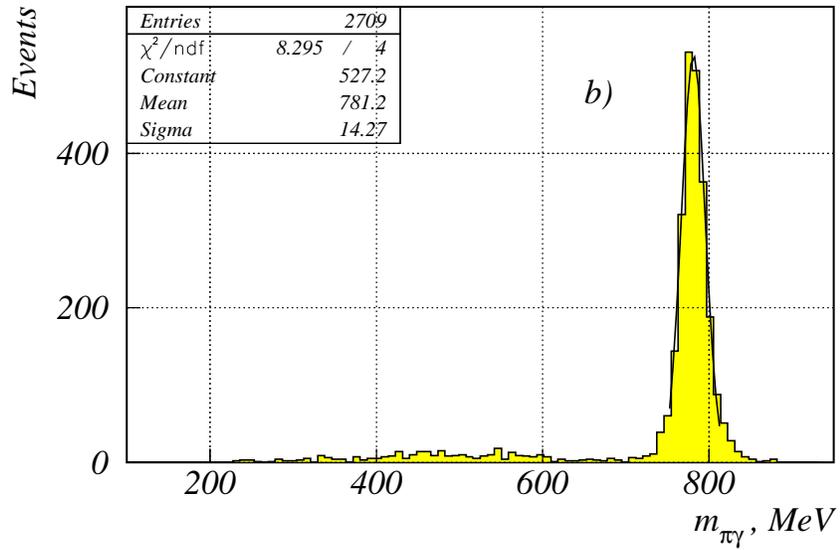,%
                   width=\textwidth}}
  \end{center}
\caption{Distribution of the events over $\pi^o \gamma$ invariant mass  
in the 
$e^+e^- \to \omega \pi^o \to \pi^o \pi^o \gamma $ process;
a --- experiment,
b --- simulation.}
\label{MPI0G}
\end{figure}

In Fig.~\ref{MPI0G} the distributions over invariant mass of
$\pi^o \gamma$- system in experimental and simulated events
are shown. It is clearly seen that most of events contain intermediate
$\omega$- meson. In Fig.~\ref{MPI0G} of two possible $\pi^o \gamma$-
pairs in an event only one with the mass closest to that
of $\omega$- meson is histogrammed. The number of experimental events
in the $\omega$- peak region is equal to 304, while expected background
from the decay $\phi \to \eta \gamma$ is 92 events. The background from
$\phi \to K_S K_L$, according to available simulation statistics, is not
more than 20 events. This background could be further suppressed by
additional constraint on invariant mass of $\pi^o \pi^o$-
system to exclude the $K_S \to 2\pi^o$ decay region.

\begin{figure}[htb]
  \begin{center}
    \mbox{\epsfig{figure=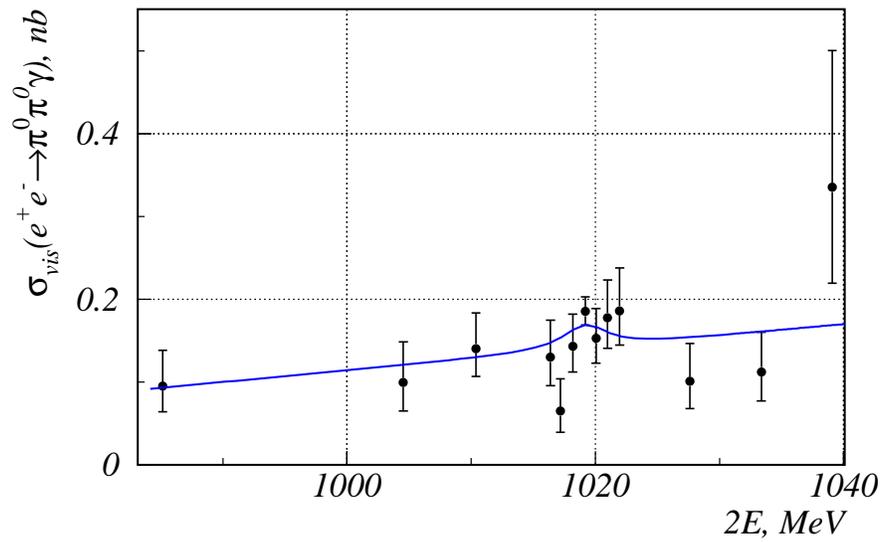,%
                  width=\textwidth}}
  \end{center}
\caption{Cross section of the $e^+e^- \to \omega \pi^o$ process.}
\label{SIGOMEGAPI0}
\end{figure}

The energy dependence of the  $e^+e^- \to \omega \pi^o$ 
cross section was approximated using parameterization (\ref{CS_INTERF})
with $B(\omega \to \pi^o \gamma)=8.5 \pm 0.5\%$, and taking into
account contributions of background processes 
$\phi \to K_S K_L, \eta \gamma$ into $ \sigma _B(E_i) $.
In this approximation due to small statistics the contribution from
$\phi \to \rho \pi , \omega \pi ^o \to \pi ^o \pi ^o \gamma$
was neglected and thus the energy dependence of the cross section
was described by a simple linear function.
As a result of fitting of the visible cross section (Fig.~\ref{SIGOMEGAPI0} )
the following results were obtained:

$\sigma _o(m_{\phi}) = 7.5 \pm 1.5 \pm 0.5~nb$ --- the cross section
of $e^+e^- \to \omega \pi ^o \to \pi ^o \pi ^o \gamma$ in the $\phi$- meson
peak  (previous ND'1991 result (\cite{ND})was $\sigma _o(m_{\phi}) = 
8.7 \pm 1.0 \pm 0.7~nb$);
$\sigma '(m_{\phi}) = 9\pm 6~MeV^{-1}$ --- relative slope of the
cross section at the same energy;
$P(\chi ^2) = 12.3/9$.

The  $g_{\rho \omega \pi }$ value  in an assumption of 
 $\rho$-meson dominance is equal to:

$g_{\rho \omega \pi }(experimental) = 18.1\pm 1.7~GeV^{-1}$.

in VDM this constant equals to:

$g_{\rho \omega \pi }(estimated) = 14.3~GeV^{-1}$.

In the ND experiment \cite{ND,ZURO1} the cross section was larger,
but it does not contradict the new SND result.
The SND data confirm conclusion of the ND experiment that pure
$\rho$(770) production cannot describe the data and to take into account 
production of higher resonances $\rho', \omega', \phi'$ is necessary.


\subsubsection{ Production of $\Delta $- barion }

When electron beam travels through the collider beam pipe the
following process may occur on nuclei of residual gas atoms:
$e^{\pm}A \to e^{\pm}\Delta \to e^{\pm}N\pi$ \cite{ACHASOV_D}.
The $\Delta (1232)$ is a barion state with quantum numbers:
$I(J^P)=\frac{3}{2}(\frac{3}{2}^+)$.
At a beam energy of $E=510~MeV$ the total cross section of this process
$\sim 3 \mu b$.  $\Delta (1232)$ decays into the following final states:
$\Delta ^+ \to p\pi^o, n\pi^+$ ט $\Delta ^o \to p\pi^-, n\pi^o$.

An attempt was taken to observe this process with SND detector and to
study it as a possible source of background at future $\phi$- factories.

The process was studied in the:
\begin{equation}
e^{\pm}p \to e^{\pm}\Delta^+ \to e^{\pm}p\pi^o
\label{ACASOV1}
\end{equation}
channel because it has a distinct signature: two photons from 
$\pi^o$- meson decay, slow proton track with high $dE/dx$ in
the drift chamber, another track of $e^{\pm}$, zero total transverse
momentum and large longitudinal momentum -- $P=510~MeV/c$.

The events with 
$N_{cp}=2$, $N_\gamma=2$, $E_{tot}/2E<0.5$, $|Z|<10~cm$, etc.
were selected. Then kinematic fit was performed.

\begin{figure}[htb]
  \begin{center}
    \mbox{\epsfig{figure=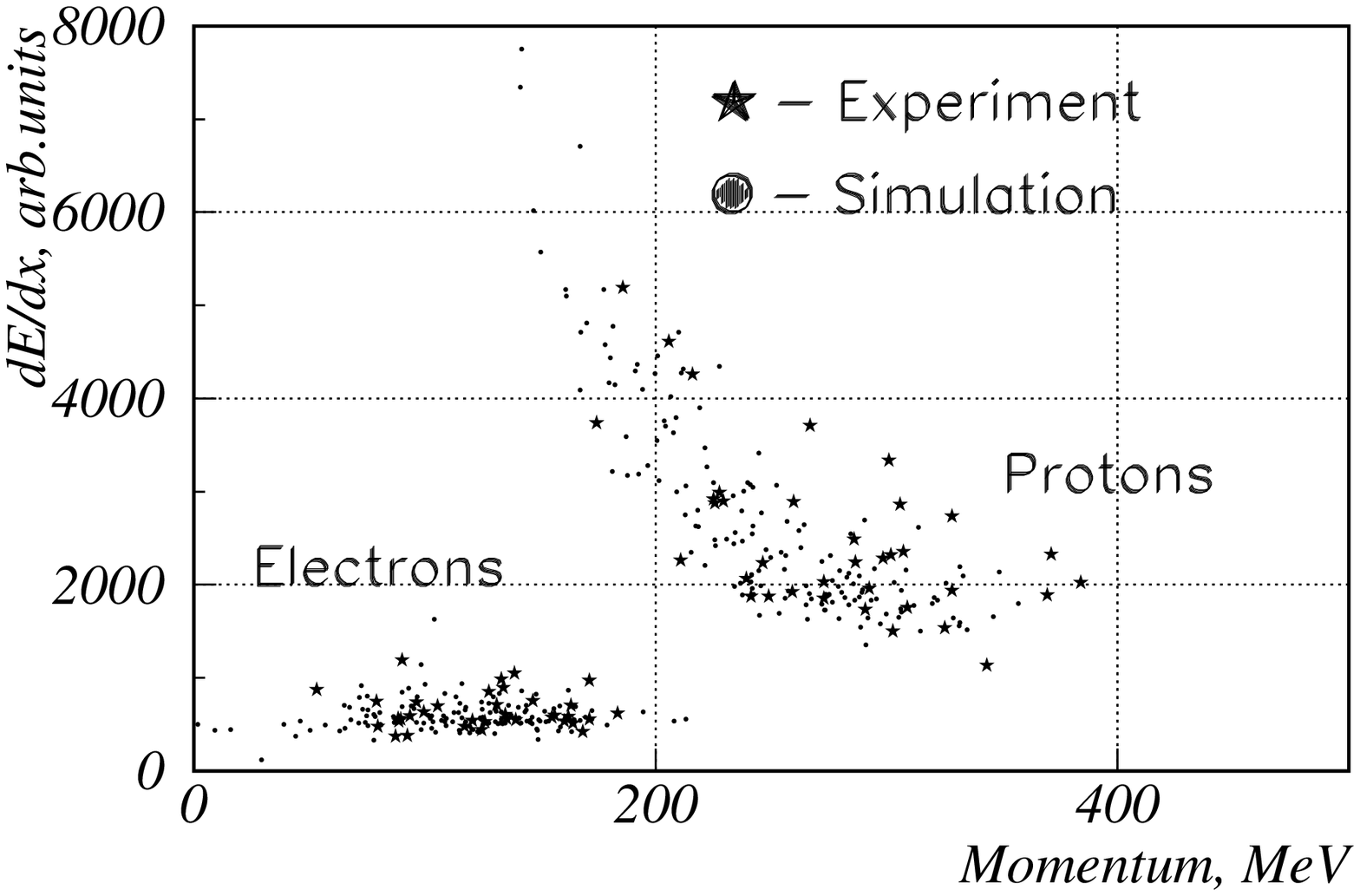,%
                  width=\textwidth}}
  \end{center} 
\caption{Dependence of $dE/dx$ in the drift chamber on particle
momentum. Experimental and simulated data for electrons and protons
from the process 
$e^{\pm}p \to e^{\pm}\Delta^+ \to e^{\pm}p\pi^o$ are shown.}
\label{ACHASOV_DEDX}

  \begin{center}
    \mbox{\epsfig{figure=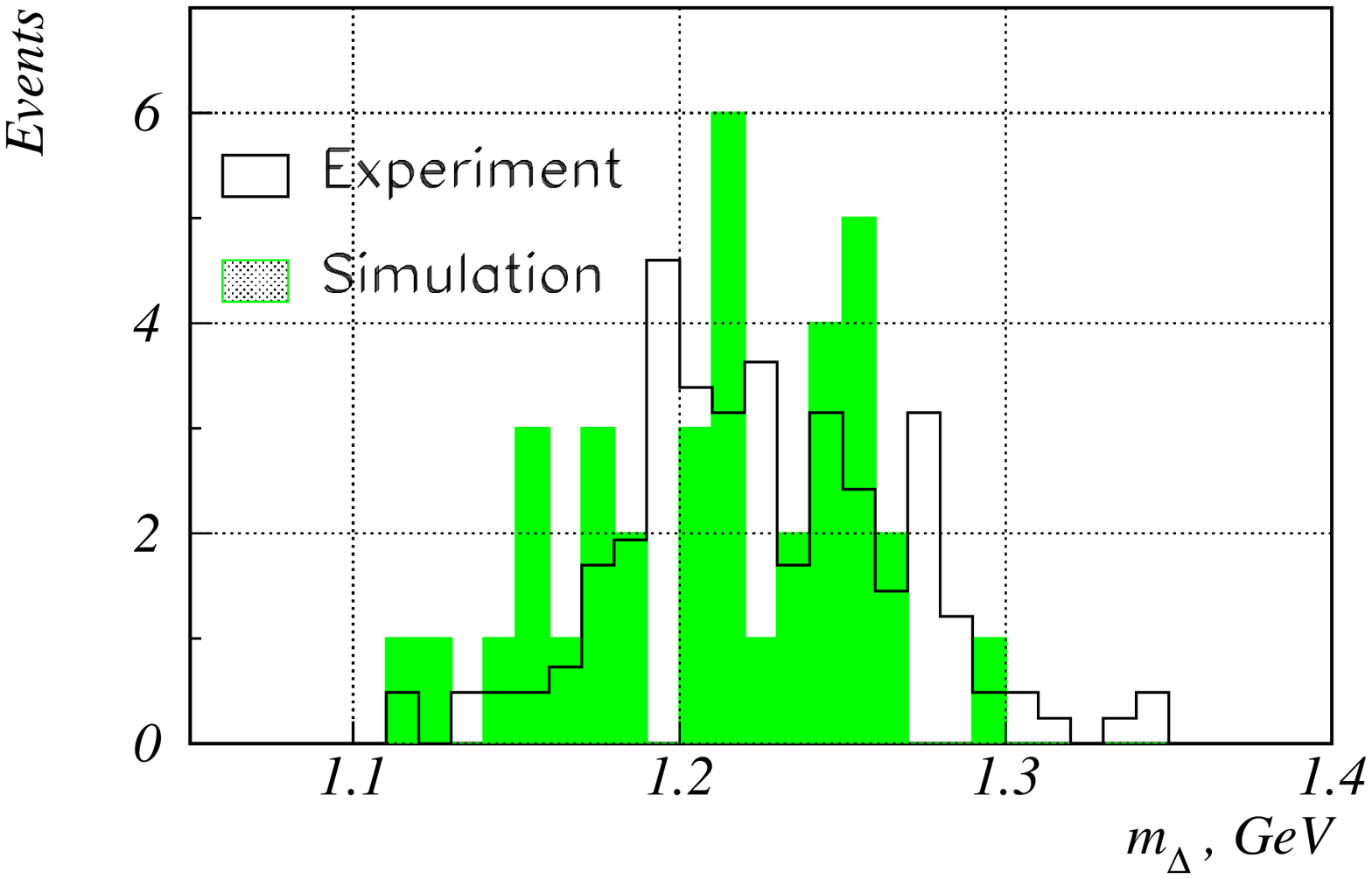,%
                  width=\textwidth}}
  \end{center} 
\caption{ 
Distribution over invariant mass of proton and pion in the
$e^{\pm}p \to e^{\pm}\Delta^+ \to e^{\pm}p\pi^o$ process}
\label{ACHASOV_D_IZOBARA}
\end{figure}

As a result of experimental events processing in the energy region
of  $\phi$- peak with integrated luminosity of
$\Delta L = 0.5~pb^{-1}$, 36 events were found. Photons in these
events have an invariant mass coinciding with that of $\pi^o$- meson
within $\pm 15~MeV$. These events are uniformly distributed in
$Z$- coordinate within $\pm 10~cm$  from the interaction point
and show no peaking there.  Among two charged particles, one with
lower specific ionization was referred to as an electron and
another -- as a proton. In Fig.~\ref{ACHASOV_DEDX} the specific
ionization in the drift chamber as a function of particle momentum
is shown for particles after kinematic fit for experimental and
simulated events. One can see, that protons are well separated
from electrons. The $\Delta$- barion mass spectrum in 
Fig.~\ref{ACHASOV_D_IZOBARA} shows good agreement between
experiment and simulation.

The background from $\phi$- meson decays is still questionable
because of insufficient simulation statistics, but it is obviously 
not high. It can be seen from absence of peaking of the events near the
beam interaction point.

Detection efficiency was estimated from the simulation and was
found to be $\varepsilon = 2\%$. Given the pressure of residual
gas of about $3\cdot 10^{-9} Torr$, mean beam current of
$15~mA$, duration of the experiment $7\cdot 10^5~s$, and taking
into account residual gas composition of
$50\%-H_2$, $30\%-CO$, $20\%-CO_2$ \cite{ACHASOV_D}, one should expect
$N\sim 4000$ events of $\Delta$- barion electroproduction at
$20 cm$ interaction region length. Taking into account the detection
efficiency, the number of observed events is expected to be equal to 30
(\ref{ACASOV1}).
It agrees well with $N_o=36$ events detected in experiment.
We should mention, that there is a considerable uncertainty in
the pressure and chemical composition of residual gas. Presence
of $CO_2$, $N_2$, etc. may lead to a considerable increase in events
number.

In conclusion, we can say, that event rate of this process
$\sim 6\cdot 10^{-3}~s^{-1}$ is too small to affect other processes
under study, but at future  $\phi$- factories it should be taken
into account (\ref{ACASOV1}) as a possible source of background.

\newpage
\section{Recent results}

In August 1997 just before printing of this preprint the following new
upper limits for processes with  multiphoton  final states were obtained:

\begin{itemize}

\item $B(\phi\to 2\pi^0) < 4 \cdot 10^{-5}$
\item $B(\phi \to \eta\pi^0) < 5 \cdot 10^{-5}$
\item $B(\phi \to 3\pi^0) < 0.8 \cdot 10^{-5}$

\end{itemize}

The event selection criteria were similar to those applied in the
studies of the processes described
above, like $\phi \to \eta\gamma \to 3\pi^0\gamma$ or $\phi \to
\eta\pi^0\gamma$.

Also was performed search for $K_S\to 3\pi^0$ decay. The event selection
criteria were adjusted for
$ \phi \to K_S K_L$ events with $K_S\to 3\pi^0$ decay and with
completely undetected $K_L$.
No indications of  $K_S\to 3\pi^0$ decay were found and the upper limit
$B(K_S\to 3\pi^0) < 1.1 \cdot 10^{-4}$ was placed.


\section{ Conclusions }

The SND detector continues data taking at VEPP--2M collider.
In 1997 the MHAD97 experiment in the energy range $2E$ from 
980 up to 1380 MeV and total integrated luminosity of
$6.3~pb^{-1}$ was carried out. The work is under way on improvement
of energy and spatial resolution of the detector. About 50\% of
experimental statistics of the FI96 experiment, collected in 1996
in the energy range $2E=984\div 1040~MeV$ were analyzed.
New preliminary results on  $\phi$- ט $\eta$- mesons decays
were obtained (Table~\ref{SUMMARY}).

\begin{table}[!htb]

\renewcommand{\arraystretch}{1.4}
\setlength{\tabcolsep}{4pt}

\caption{ SND preliminary results on 
$\phi (1020), \eta(550), K_S$ decays, compared with CMD-2
\cite{CMD2_PROCEED}, \cite{CMD2_1}, \cite{CMD2_2} results and data
from PDG \cite{PDG}.}
\label{SUMMARY}
\footnotesize
\begin{center}

\begin{tabular}{|l|c|c|c|}

\hline Decay mode& SND & CMD-2 & PDG(1996) \\

\hline 
$\phi \to \eta \gamma$ & 
$(1.30 \pm  0.06 \pm 0.07)\%$ & 
$(1.25 \!\pm\! 0.10) \%$ &
$(1.26 \!\pm\! 0.06)\%$ \\

\hspace{5mm} $\eta\to\gamma\gamma$ & $(1.37\pm 0.04\pm 0.08)\%$ & & \\
\hspace{5mm} $\eta\to 3\pi ^0$ & $(1.25\pm 0.03\pm 0.08)\%$ & & \\
\hspace{5mm} $\eta\to \pi ^+ \pi ^- \pi ^0$ & $(1.04 \pm 0.08 \pm 0.08)\%$ & & \\

\hline 
$\phi \to \pi ^0 \gamma$ & 
$(0.13 \pm 0.005\pm 0.008)\%$ & 
--- &
$(0.131 \!\pm\! 0.013)\%$ \\

\hline 
 $\phi \to \eta' (958) \gamma$ & 
 $<1.7\!\cdot\! 10^{-4}$ & 
$<2\!\cdot\! 10^{-4}$ &
$<4.1\!\cdot\! 10^{-4}$ \\

\hline  
$\phi \to K_S K_S \gamma$ &
$<3.2\!\cdot\!10^{-4}$ & 
--- & 
--- \\

\hline 
$\phi \to \pi^0 \pi^0 \gamma$ &
$(1.1\!\pm\! 0.2)\!\cdot\! 10^{-4}$ &
--- &
$<1\!\cdot\! 10^{-3}$ \\

\hline 
$\phi\to\pi^+\pi^-\gamma$ & 
 --- & 
$<1.5\!\cdot\! 10^{-5}$ &
$<7\!\cdot\! 10^{-3}$ \\

\hline 
$\phi\to\mu^+\mu^-\gamma$ & 
--- & 
$ (2.3\!\pm\! 1.0)\!\cdot\! 10^{-5}$ &
--- \\

\hline 
$\phi \to \eta \pi^0 \gamma$ & 
$(1.3\pm 0.5)\!\cdot\! 10^{-4}$ &
--- &
$<2.5\!\cdot\! 10^{-3}$ \\

\hline 
$\phi\to f_0 \gamma$ & 
$(4.7\!\pm\! 1.0)\!\cdot\! 10^{-4}$ & 
$<1-7\!\cdot\! 10^{-4}$ &
$<2\!\cdot\! 10^{-3}$ \\

\hline 
$\phi \to \rho \gamma$ &
--- &
$<3\!\cdot\! 10^{-4}$ & 
$<7\!\cdot\! 10^{-2}$ \\

\hline 
$\phi \to K_S K_L$ & 
$0.31\!\pm\! 0.02$ &
$0.325\!\pm\! 0.010$ & 
$0.341\!\pm\! 0.005$ \\

\hline 
$K_S\to 3\pi^0$ & 
 $<1.1\!\cdot\! 10^{-4}$&
--- &
$<3.7\!\cdot\!10^{-5}$\\

\hline 
$\phi \to \mu^+ \mu^-$ & 
$(2.4\!\pm\! 1.0)\!\cdot\! 10^{-4}$ &
--- & 
$(2.48\!\pm\! 0.34)\!\cdot\! 10^{-4}$ \\

\hline 
$\phi \to e^+ e^-$ & 
--- & 
$(2.84\!\pm\! 0.06)\!\cdot\! 10^{-4}$ &
{$(3.00\!\pm\! 0.06)\!\cdot\! 10^{-4}$ }\\

\hline 
$\phi \to \eta e^+ e^-$ &
--- & 
$(1.1\!\pm\! 0.5)\!\cdot\! 10^{-4}$ & 
$(1.3_{-0.6}^{+0.8})\!\cdot\! 10^{-4}$ \\

\hline 
$\phi \to \pi^0 e^+ e^-$ &
--- & 
$(1.1\!\pm\! 0.8)\!\cdot\! 10^{-5}$ &
$<1.2\!\cdot\! 10^{-4}$ \\

\hline  
\begin{minipage}{25mm}
$\phi \to \omega \pi$ \\ $(\pi^+\pi^-\pi^0\pi^0)$ 
\end{minipage}
& 
$<7\!\cdot\! 10^{-5}$ &
--- & 
--- \\

\hline 
$\phi\hspace*{-1.5mm}\to\hspace*{-1.5mm}\pi^+\pi^-\pi^+\pi^-$ &
--- & 
$<3\!\cdot\! 10^{-5}$ &
$<8.7\!\cdot\! 10^{-4}$ \\

\hline 
$\phi\to 2\pi^0$ &
$<4\!\cdot\! 10^{-5}$ &
--- &
--- \\

\hline 
$\phi\to 3\pi^0$ &
$<0.8\!\cdot\! 10^{-5}$ &
--- &
--- \\

\hline 
$\phi\to \eta\pi^0$ &
$<7\!\cdot\! 10^{-5}$ &
--- &
--- \\

\hline 
$\eta \to e^+e^-$ &
$<9\!\cdot\! 10^{-4}$ & 
--- & 
$<3\!\cdot\! 10^{-4}$ \\

\hline 
$\eta \to \pi^0 \pi^0$ & 
$<7\!\cdot\! 10^{-4}$ &
--- & 
--- \\

\hline 
$\eta \to \pi^+ \pi^-$ & 
--- & 
$<9\!\cdot\! 10^{-4}$ & 
$<1.5\!\cdot\! 10^{-3}$ \\

\hline
\end{tabular}
\end{center}

\vspace{3cm}

\end{table}

Cross sections of the following nonresonant processes were
measured in this energy region:

$\sigma _o(e^+e^-\to \omega \pi^o (\omega \to \pi^+ \pi^- \pi^o)) =
                   6.4\pm 0.5 \pm 1.0~nb$;

$\sigma _o(e^+e^-\to \omega \pi^o (\omega \to \pi^o \gamma)) =
                   7.5\pm 1.5 \pm 0.5~nb$;

$\sigma _o(e^+e^-\to \mu^+\mu^-) = 96.8\pm 0.9 \pm 5.0~nb$;

$\sigma _o(e^+e^-\to \gamma \gamma \gamma) = 1.56\pm 0.16 \pm 0.20~nb$,
 at  $\vartheta _{min} > 27^o$;

$\sigma _o(e^+e^-\to e^+e^- \gamma) = 550\pm 10 \pm 55 nb$,
 at  $\vartheta _{min} > 27^o$ ט $E_{\gamma}>5~MeV$.

Electroproduction of  $\Delta$- barion on residual gas nuclei was seen:
$e^{\pm}A \to e^{\pm}\Delta \to e^{\pm}N\pi$.


\newpage

\clearpage

\clearpage

\clearpage

\newpage

\end{document}